\documentclass[prd,aps,twocolumn,preprintnumbers, showpacs,nofootinbib,superscriptaddress,notitlepage,floatfix]{revtex4-1}
\usepackage{amsthm,amsmath,amssymb}
\usepackage{xcolor}
\usepackage{graphicx}
\usepackage{multirow}
\usepackage[colorlinks, citecolor=blue,anchorcolor=blue,menucolor=blue, linkcolor=blue,filecolor=blue,runcolor=blue,urlcolor=blue,frenchlinks=true]{hyperref}

\DeclareMathOperator{\Tr}{Tr}

\begin{document}

\title{Nonperturbatively Renormalized Nucleon Gluon Momentum Fraction in the Continuum Limit of $N_f=2+1+1$ Lattice QCD}
\author{Zhouyou Fan}
\affiliation{Department of Physics and Astronomy, Michigan State University, East Lansing, MI 48824}

\author{Huey-Wen Lin}
\affiliation{Department of Physics and Astronomy, Michigan State University, East Lansing, MI 48824}
\affiliation{Department of Computational Mathematics,
  Science and Engineering, Michigan State University, East Lansing, MI 48824}

\author{Matthew Zeilbeck}
\affiliation{Department of Physics and Astronomy, Michigan State University, East Lansing, MI 48824}

\preprint{MSUHEP-22-028}

%%%%%%%%%%%%%%%%%%%%%%%%%%%%%%%%%%%%%%%%%%%%%%%%%%%%%%%%%%%%%%%%%%%%%%%%%%%%%%%%
\begin{abstract}
We present the nonperturbatively renormalized nucleon gluon momentum fraction using ensembles with $2+1+1$ flavors of highly improved staggered quarks (HISQ), generated by MILC Collaboration.
The calculation is done using clover fermions for the valence action with three pion masses, $220$, $310$ and $690$~MeV and three lattice spacings, 0.09, 0.12, and 0.15~fm.
The renormalization is done using RI/MOM nonperturbative renormalization and using cluster-decomposition error reduction (CDER) to enhance the signal-to-noise ratio of the renormalization constant.
We find the CDER technique is particularly important to  improve the signal at the finer lattice ensembles where the lattice volume is larger.
We extrapolate the gluon momentum fraction to the continuum-physical limit and obtain $\langle x \rangle_g = 0.492(52)_\text{stat.+NPR}(49)_\text{mixing}$ in the $\overline{\text{MS}}$ scheme at 2~GeV, where first error includes the statistical error and uncertainties in nonperturbative renormalization, while the latter systematic accounts for ignoring quark mixing.
Our gluon momentum fraction is consistent with other recent lattice-QCD results at physical pion mass.
\end{abstract}
\maketitle

%%%%%%%%%%%%%%%%%%%%%%%%%%%%%%%%%%%%%%%%%%%%%%%%%%%%%%%%%%%%%%%%%%%%%%%%%%%%%%%%
\section{Introduction}

The gluon momentum fraction $\langle x \rangle_g$ of the nucleon is important to particle and nuclear physics.
It can be measured as the momentum fraction carried by gluons in the infinite momentum frame and must satisfy the momentum sum rule $\langle x \rangle_g + \langle x \rangle_q=1$ with the sum of the quark momentum fraction.
These momentum fractions are key inputs to understanding the proton mass and spin decomposition, which are major outstanding questions in hadronic physics.
The gluon momentum fraction is connected to the unpolarized nucleon gluon parton distribution function (PDF) $g(x)$ via
\begin{equation}
\langle x \rangle_g =\int^1_0 dx\ x g(x).
\label{eq:xg-moment}
\end{equation}
The gluon PDF is an important input to many theory predictions used in the hadron colliders~\cite{Dulat:2015mca,Harland-Lang:2014zoa,Ball:2017nwa,Alekhin:2017kpj,Accardi:2016qay,Harland-Lang:2019pla,Bertone:2017bme,Manohar:2017eqh}.
For example, $g(x)$ needs to be known precisely to calculate the cross section for processes in $pp$ collisions, including the cross section for Higgs-boson production and jet production at the Large Hadron Collider (LHC)~\cite{CMS:2012nga,Kogler:2018hem}.
Ongoing and future experiments, such as new experiments at the Jefferson Lab 12-GeV facility and the U.S.-based Electron-Ion Collider (EIC)~\cite{Accardi:2012qut}, planned to be built at Brookhaven National Lab, will further our knowledge of the gluon PDF~\cite{Arrington:2021biu,Aguilar:2019teb,AbdulKhalek:2021gbh}.

Lattice quantum chromodynamics (lattice QCD or LQCD) is a theoretical method that can provide full systematic control in calculating QCD quantities in the nonperturbative regime and can provide useful information for improving our knowledge of the gluon structure of the nucleon, independent from experiments.
There have been many lattice calculations of the nucleon quark momentum fraction $\langle x \rangle_q $ (see reviews in Refs.~\cite{Lin:2017snn,Constantinou:2020hdm}), but still relatively few attempts for the gluon counterpart~\cite{Gockeler:1996zg,Liu:2011vcs,QCDSF:2012mkm,Deka:2013zha}.
This is mainly due to the fact that any gluon observable on the lattice is extremely noisy.
Furthermore, the renormalization for even the gluon-only momentum fraction has been difficult to calculate nonperturbatively (at large volumes). 
Early lattice-QCD studies calculated $\langle x \rangle_g$ of the nucleon on quenched lattices using heavy pion masses and gave $\langle x \rangle_g\in[0.3,0.6]$~\cite{Gockeler:1996zg,Liu:2011vcs,QCDSF:2012mkm,Deka:2013zha}. 
There have been a number of dynamical calculations of the gluon momentum fraction of the nucleon using 2-flavor (degenerate up and down sea quarks), 2+1-flavor (including strange quark) and 2+1+1-flavor (including charm quark) lattice calculations by ETMC, $\chi$QCD, and MIT lattice groups~\cite{Alexandrou:2016ekb,Alexandrou:2017oeh,Shanahan:2018pib,Yang:2018bft,Alexandrou:2020sml,Alexandrou:2020sml}; see the summary in Table~\ref{tab:latticemoments}.
Additional smearing and large numbers of statistical measurements are typically needed to produce a usable gluon signal.
Since 2018, $\chi$QCD and MIT lattice groups have used nonperturbative renormalization on the gluon operators.
There has been an attempt by $\chi$QCD to study the lattice-spacing dependence using $2+1$-flavor ensembles using partially quenched mixed actions (where the valence pion masses are allowed to be different from the sea pion masses).
Although progress has been made in recent years, there is still disagreement between the lattice determination of the gluon momentum fraction and those obtained from taking the integral of the global-fit gluon PDF determination in Eq.~\ref{eq:xg-moment}.
Reference~\cite{Lin:2017snn} quoted numbers from multiple gluon PDF determinations (NNPDF3.1, CT14, MMHT14, ABMP16, CJ15, and HERAPDF2.0), yielding a weighted-average gluon momentum fraction of 0.411(8). 
Since then, JAM19 and CT18 have published updated values, 0.403(2) and 0.413(8), respectively.
More lattice studies are needed to understand the potential discrepancy between lattice calculations and global-fit results.

\begin{table*}[htbp!]
\centering
\begin{tabular}{|c|c|c|c|c|c|c|c|c|}
\hline
  Group & $N_f$ &  $a$ (fm) & $M_\pi^\text{val}$ (MeV) & Fermion & $N_\text{meas}$  &  Renorm. &  G-smearing & $\langle x \rangle_g $\\
\hline
\hline
 ETMC16~\cite{Alexandrou:2016ekb} & 2+1+1 & 0.08 &$370$   & TM &  34,470     & 1-loop & 2-stout & $0.284(27)_\text{stat.}(17)_\text{ES}(24)_\text{PT}$\\
\hline
 ETMC16~\cite{Alexandrou:2016ekb} & 2 & 0.09 &$131$   & TM &  209,400     & 1-loop & 2-stout & $0.267(22)_\text{stat.}(19)_\text{ES}(24)_\text{PT}$\\
\hline
 ETMC17~\cite{Alexandrou:2017oeh} & 2 & 0.09 &$131$   & TM  &  209,400     & 1-loop & 2-stout & $0.267(12)_\text{stat.}(10)_\text{ES}$\\
 \hline
 MIT18~\cite{Shanahan:2018pib} & $2+1$ &  0.12 &  $450$ &  clover  & 572,663 & RI-MOM & Wilson flow & $0.54(8)_\text{stat.}$\\
 \hline
 $\chi$QCD18a~\cite{Yang:2018bft} &
 2+1 &
 0.114 &
 [135, 372]\footnote{partially quenched calculation on domain-wall fermion  $M_\pi^\text{sea} = 140$-MeV lattice} &
 overlap & 81 cfgs
 & RI-MOM &
  1-HYP &
 $0.47(4)_\text{stat.}(11)_{\text{NPR+mixing}}$ \\
 \hline
 $\chi$QCD18b~\cite{Yang:2018nqn} &
  2+1 & $[0.08,0.14]$ & [140,400] & overlap & $[81,309]$ cfgs & RI-MOM & 1-HYP & $0.482(69)_\text{stat.}(48)_\text{cont.}$
 \\
   \hline
 ETMC20~\cite{Alexandrou:2020sml} & $2+1+1$ & 0.08 & $139.3$ & TM & 48,000    & 1-loop & 10-stout & $0.427(92)_\text{stat.}$ \\
\hline
$\chi$QCD21~\cite{Wang:2021vqy} & $2+1$ &  0.14 &
[171, 391]\footnote{partially quenched calculation on domain-wall fermion $M_\pi^\text{sea} = 171$-MeV lattice} &
overlap &
 8,200 & RI-MOM &
 1-HYP &
$0.509(20)_\text{stat.}(23)_\text{cont.} $\\
\hline
MSULat22 & $2+1+1$ &  [0.09,0.15]  &
[220,700]\footnote{clover-on-HISQ mixed action with valence pion masses tuned to lightest sea-quark ones} &
clover &
$10^5$--$10^6$ & RI-MOM &
 5-HYP &
$0.492(52)_\text{stat.+NPR}(49)_\text{mixing}$ \\
(this work) & &  & &  & & &  &\\
\hline
\end{tabular}
\caption{
Summary of lattice dynamical calculations of the nucleon gluon moment sorted by year.
The columns from left to right show for each calculation:
the number of flavors of quarks in the QCD vacuum ($N_f$),
the lattice spacing ($a$) in fm,
the valence pion mass ($M_\pi^\text{val}$) in MeV,
the valence fermion action (``Fermion''), where ``TM'' stands for twisted-mass fermion action,
the number of measurements of the nucleon correlators ($N_\text{meas}$),
the renormalization method (``Renorm.'') indicating 1-loop perturbative calculations or RI-MOM nonperturbative renormalization,
the smearing technique used to improve the gluon signals (``G-smearing''),
and the obtained gluon momentum fraction ($\langle x \rangle_g$) renormalized at 2-GeV scale in $\overline{\text{MS}}$ scheme.
The lattice errors coming from different sources are marked as
``stat.'' for statistical, ``cont.'' for continuum-extrapolation (or lack thereof), ``ES'' for excited state contamination (but later calculations remove them, folding this error into the statistical), ``PT'' for perturbative renormalization, ``NPR'' for nonperturbative renormalization, and ``mixing'' for the mixing with the quark sector.
\label{tab:latticemoments}
}
\end{table*}

The gluon momentum fraction remains an important calculation target despite recent developments in pseudo-PDF~\cite{Balitsky:2019krf} and quasi-PDF~\cite{Zhang:2018diq,Wang:2019tgg} approaches, which have opened up opportunities to calculate the full $x$-dependence of the gluon PDF.
The first attempt to determine the nucleon gluon PDF in a lattice-QCD calculation was done based on a quasi-PDF approach~\cite{Fan:2018dxu}, but it did not obtain a sufficient signal to reconstruct the gluon PDF $g(x)$.
Lattice calculations to access the nucleon, pion, and kaon gluon PDFs $g(x)$ followed~\cite{Fan:2020cpa,Fan:2021bcr,HadStruc:2021wmh,Salas-Chavira:2021wui} using the pseudo-PDF approach.
However, the calculation of the gluon PDF via pseudo-PDF method gives the ratio of $xg(x)/\langle x \rangle_g$, and one still needs a direct lattice calculation of $\langle x \rangle_g$ to extract the gluon PDF by itself.
Therefore, the lattice gluon momentum fraction remains an important input in the era of $x$-dependent PDF lattice hadronic calculations. 

In this work, we present a lattice-QCD calculation of gluon momentum fraction $\langle x \rangle_g$ in the physical-continuum limit using clover fermions on  $N_f=2+1+1$ HISQ lattices with three lattice spacings, 0.09, 0.12, and 0.15~fm, and three pion masses, 690, 310, and 220~MeV.
The rest of the paper is organized as follows.
In Sec.~\ref{sec:lattice-ME}, we present the lattice setup and examples of how we extract the ground-state matrix elements from the lattice correlators to obtain the bare gluon momentum fraction of the nucleon.
In Sec.~\ref{sec:NPR}, the method and results of the nonperturbative renormalization of the gluon momentum fraction are discussed.
In Sec.~\ref{sec:results}, we extrapolate the renormalized gluon momentum fractions of different ensembles to the physical pion mass and continuum limit, then compare our results with other lattice calculations and global fits.
We discuss possible systematics that may contribute to additional uncertainties in our results. 
A summary and the outlook for future calculations of the nucleon gluon momentum fraction can be found in Sec.~\ref{sec:conclusion}.

%%%%%%%%%%%%%%%%%%%%%%%%%%%%%%%%%%%%%%%%%%%%%%%%%%%%%%%%%%%%%%%%%%%%%%%%%%%%%%%%
\section{Lattice Setup and Bare Gluon Matrix Elements}\label{sec:lattice-ME}
%%%%%%%%%%%%%%%%%%%%%%%%%%%%%%%%%%%%%%%%%%%%%%%%%%%%%%%%%%%%%%%%%%%%%%%%%%%%%%%%

We present our calculation of the nucleon gluon PDFs using clover valence fermions on four ensembles with $N_f = 2+1+1$ highly improved staggered quarks (HISQ)~\cite{Follana:2006rc} generated by MILC Collaboration~\cite{Bazavov:2012xda} with three different lattice spacings ($a\approx 0.9, 0.12$ and 0.15~fm) and three pion masses (220, 310, and 690~MeV), as shown in Table~\ref{table-data}.
The masses of the clover quarks are tuned to reproduce the lightest light and strange sea pseudoscalar meson masses done by PNDME Collaboration~\cite{Rajan:2017lxk,Bhattacharya:2015wna,Bhattacharya:2015esa,Bhattacharya:2013ehc}. 
PNDME calculated the nucleon quark isovector, helicity and transversity moments using the clover-on-HISQ ensembles (``mixed action'') in Ref.~\cite{Mondal:2020cmt};
the quark momentum fraction results obtained in Ref.~\cite{Mondal:2020cmt} are consistent with the phenomenological global-fit values. 
In this work, we use five HYP-smearing~\cite{Hasenfratz:2001hp} steps on the gluon loops to reduce the statistical uncertainties, based on the study in Ref.~\cite{Fan:2018dxu}.
Table~\ref{table-data} shows the ensemble information, such as the lattice size $L^3\times T$ and number of total two-point correlator measurements $N_\text{meas}$ in this calculation. 
The number of measurements varies $10^5$--$10^6$ for different ensembles. 

\begin{table}[!htbp]
\centering
\begin{tabular}{|c|c|c|c|c|}
\hline
  ensemble & a09m310 & a12m220 & a12m310 & a15m310 \\
\hline
  $a$ (fm) & $0.0888(8)$ & $0.1184(10)$ & $0.1207(11)$  & $0.1510(20)$ \\
\hline
  $L^3\times T$ & $32^3\times 96$ & $32^3\times 64$ & $24^3\times 64$ & $16^3\times 48$ \\
\hline
  $M_{\pi}^\text{val}$ (MeV) & $313.1(13)$ & $226.6(3)$ & $309.0(11)$ & $319.1(31)$\\
\hline
  $M_{\eta_s}^\text{val}$ (MeV) & $698.0(7)$
  &  N/A  
  & $684.1(6)$ & $687.3(13)$\\
\hline
  $N_\text{cfg}$ & 1009 & 957 & 1013 & 900   \\
\hline
  $N_\text{meas}$ & 387,456 &  1,466,944  & 324,160 & 259,200   \\
  $t_\text{sep}$ & [8,12]& [7,11]   & [7,11] & [5,9]  \\
\hline
\end{tabular}
\caption{ 
Lattice spacing $a$, valence pion mass $M_\pi^\text{val}$ and $\eta_s$ mass $M_{\eta_s}^\text{val}$, lattice size $L^3\times T$, number of configurations $N_\text{cfg}$, number of total two-point correlator measurements $N_\text{meas}^\text{2pt}$, and source-sink separation $t_\text{sep}$ used in the three-point correlator fits of $N_f=2+1+1$ clover valence fermions on HISQ ensembles generated by the MILC Collaboration and analyzed in this study.
}
\label{table-data}
\end{table}

On the lattice, we calculate the two-point correlator for a nucleon $N$ via
\begin{equation}
C_N^\text{2pt}(P_z;t) =
 \langle 0|\Gamma\int d^3y\, e^{-iy_z P_z}\chi(\vec y,t)\chi(\vec 0,0)|0\rangle, \label{eq:2ptdef}
\end{equation} 
where $P_z$ is the boosted nucleon momentum along the spatial $z$-direction,
$t$ is lattice Euclidean time, $\chi(y)=\epsilon^{lmn}[{u(y)^l}^Ti\gamma_4\gamma_2\gamma_5 d^m(y)]u^n(y)$ (where $\{l,m,n\}$ are color indices, $u(y)$ and $d(y)$ are the quark operators) is the nucleon interpolation operator, and $\Gamma=\frac{1}{2}(1+\gamma_4)$ is the projection operator.
We also calculate the three-point correlator to obtain the matrix elements needed to extract the gluon momentum fraction via
\begin{equation}
C_N^\text{3pt}(P_z;t_\text{sep},t)
= \int d^3y\, e^{-iy_z P_z}\langle \chi(\vec y,t_\text{sep})| O_{g,tt}(t)|\chi(\vec 0,0)\rangle,\label{eq:3ptdef}
\end{equation}
where $t_\text{sep}$ is the source-sink separation and $t$ is the gluon-operator insertion time. 
The operator for the gluon momentum fraction $O_{g,tt}(t)$ is 
\begin{equation}
O_{g,\mu\nu} \equiv \sum_{i=x,y,z,t} F^{\mu i} F^{\nu i}
      - \frac{1}{4} \sum_{i,j=x,y,z,t} F^{ij} F^{ij}, 
\label{eq:op_def}
\end{equation}
where the field tensor $F_{\mu\nu}$ is
\begin{equation}
F_{\mu\nu} = \frac{i}{8a^2g} (\mathcal{P}_{[\mu,\nu]} + \mathcal{P}_{[\nu,-\mu]} + \mathcal{P}_{[-\mu,-\nu]} + \mathcal{P}_{[-\nu,\mu]}),
\end{equation}
with the plaquette $\mathcal{P_{\mu,\nu}} = U_{\mu}(x)U_{\nu}(x+a\hat{\mu})U^{\dag}_{\mu}(x+a\hat{\nu})U^{\dag}_{\nu}(x)$ and $\mathcal{P_{[\mu,\nu]}} = \mathcal{P}_{\mu,\nu} - \mathcal{P}_{\nu,\mu}$.
The same gluon operator was also used in the recent calculation of the gluon momentum fraction by $\chi$QCD, ETMC and MIT lattice collaborations~\cite{Alexandrou:2016ekb,Alexandrou:2017oeh,Shanahan:2018pib,Yang:2018bft,Alexandrou:2020sml,Alexandrou:2020sml,Yang:2018nqn}. 

Using the two- and three-point correlators, we can extract the ground-state nucleon matrix elements that lead to the gluon momentum fraction.
We use Gaussian momentum smearing for the quark fields~\cite{Bali:2016lva} $q(x)+\alpha\sum_{j}U_j(x)e^{i(\frac{2\pi}{L})\textbf{k}\hat{e}_{j}}q(x+\hat{e}_{j})$, so that we can calculate the momentum fraction using $P_z \neq 0$;
these correlators have been neglected in previous calculations due to their worse signal-to-noise ratios relative to those obtained from $P_z = 0$.
We fit the two-point and three-point correlators to the energy-eigenstate expansion
\begin{align}
\label{eq:2ptC}
&C_N^\text{2pt}(P_z;t) \nonumber \\
&= |A_{N,0}|e^{-E_{N,0}t}+ |A_{N,1}|e^{-E_{N,1}t}+ \ldots,
\end{align}
\begin{align}
\label{eq:3ptC}
&C_N^\text{3pt}(z,P_z;t_\text{sep},t) \nonumber \\
&= |A_{N,0}|^2\langle 0| O_{g,tt}|0\rangle e^{-E_{N,0}t_\text{sep}} \nonumber \\
&+ |A_{N,0}||A_{N,1}|\langle 0|{\cal O}|1\rangle e^{-E_{N,1}(t_\text{sep}-t)}e^{-E_{N,0}t} \nonumber \\
&+ |A_{N,0}||A_{N,1}|\langle 1|{\cal O}|0\rangle e^{-E_{N,0}(t_\text{sep}-t)}e^{-E_{N,1}t} \nonumber \\
&+ |A_{N,1}|^2\langle 1|{\cal O}|1\rangle e^{-E_{N,1}t_\text{sep}}+ \ldots,
\end{align}
where the ground (first-excited) state amplitudes and energies $A_{N,0}$, $E_{N,0}$ ($A_{N,1}$, $E_{N,1}$), are obtained from the two-state fits of the two-point correlators.
The parameters $\langle 0|{\cal O}|0\rangle$, $\langle 0|{\cal O}|1\rangle$ ($\langle 1|{\cal O}|0\rangle=\langle 0|{\cal O}|1\rangle^*$), and $\langle 1|{\cal O}|1\rangle$ are the ground-state, the ground--excited-state, and the excited-state matrix elements, respectively.
The matrix elements can be extracted by using the two-state simultaneous fits (``two-sim fits'') of the three-point correlators using multiple $t_\text{sep}$ inputs.

\begin{figure*}[htbp]
\centering
\includegraphics[width=0.45\textwidth]{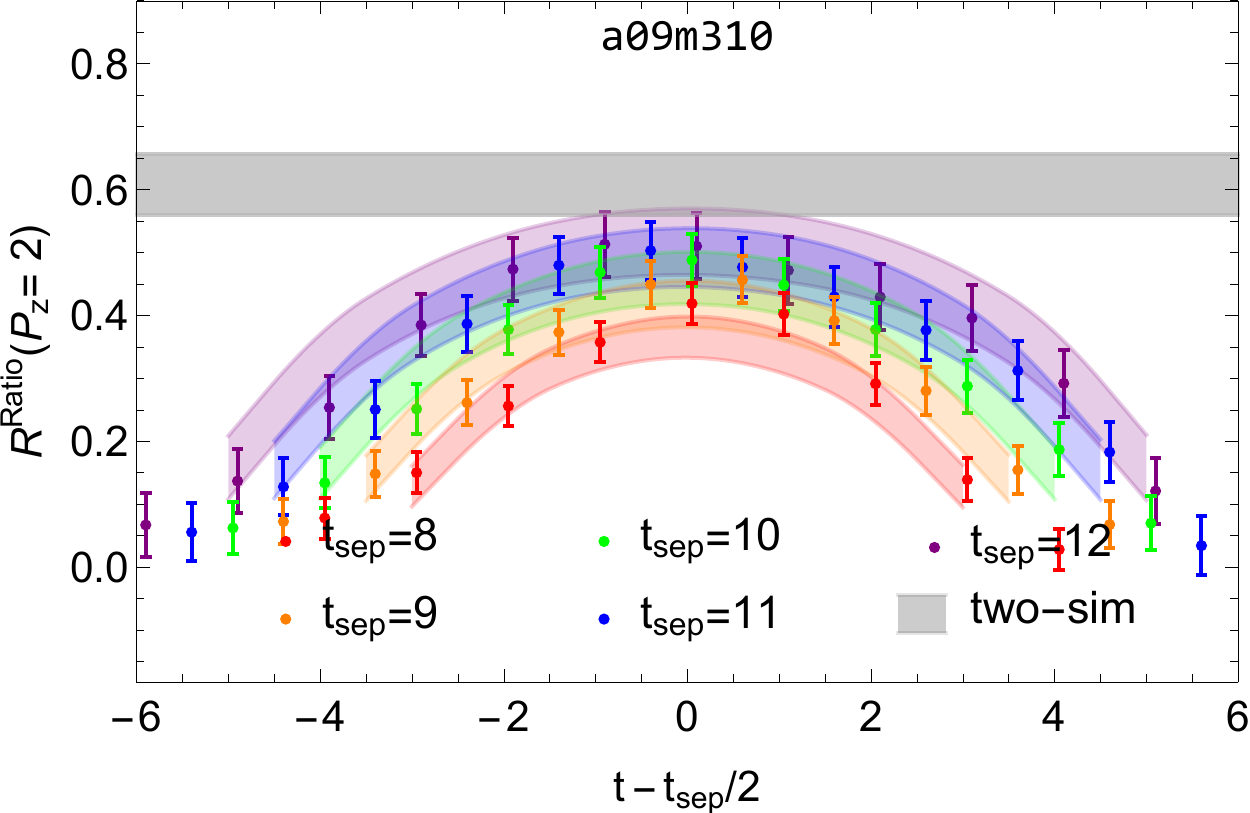}
\centering
\includegraphics[width=0.265\textwidth]{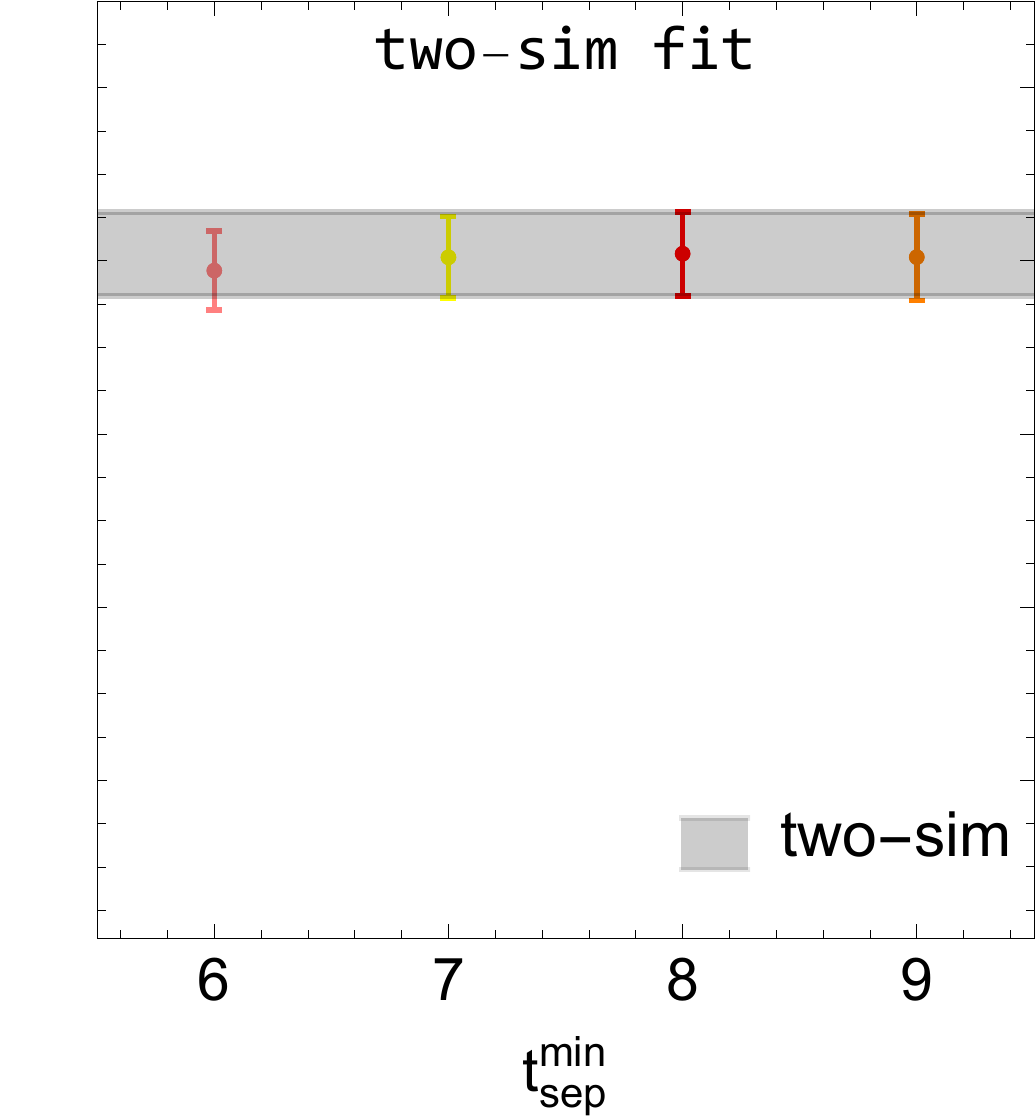}
\includegraphics[width=0.268\textwidth]{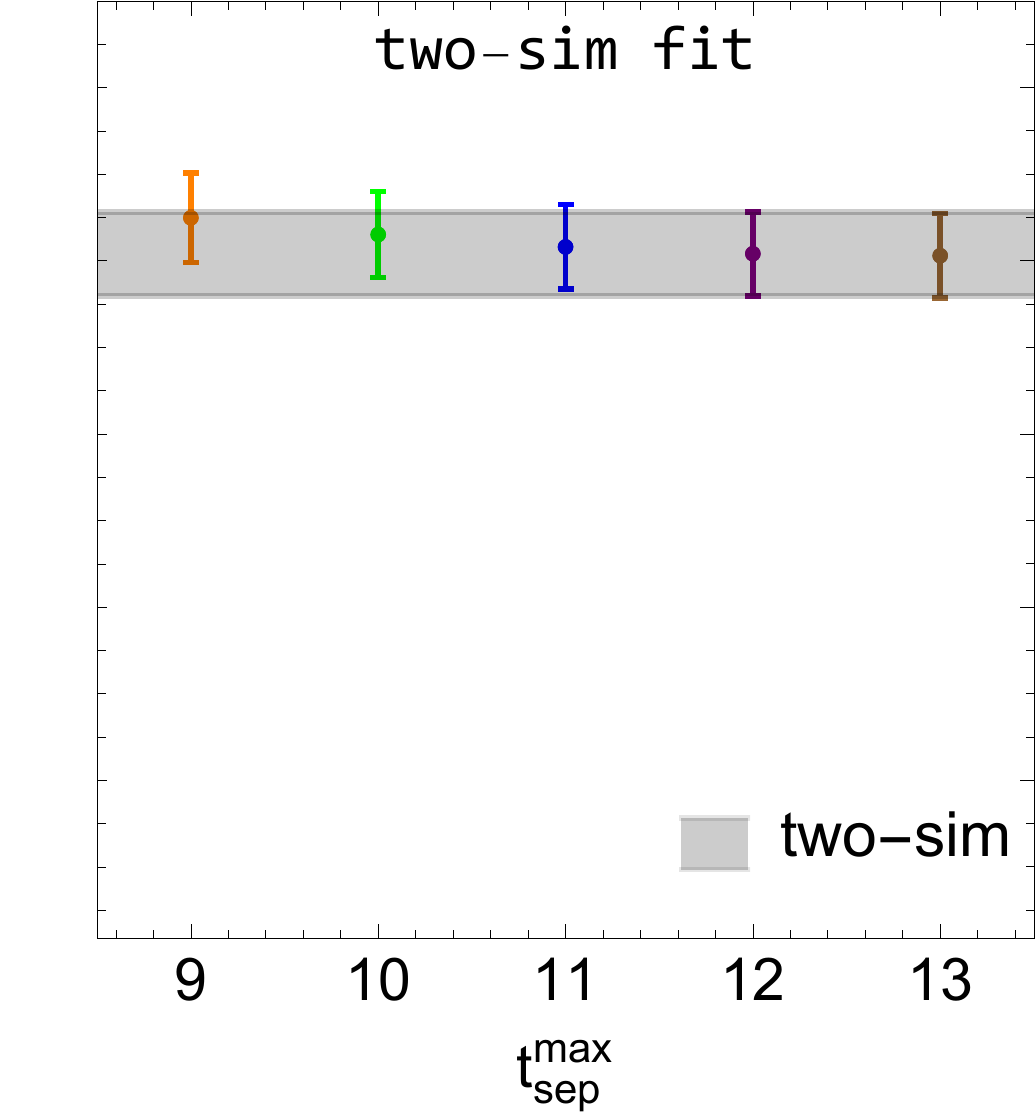}
\includegraphics[width=0.45\textwidth]{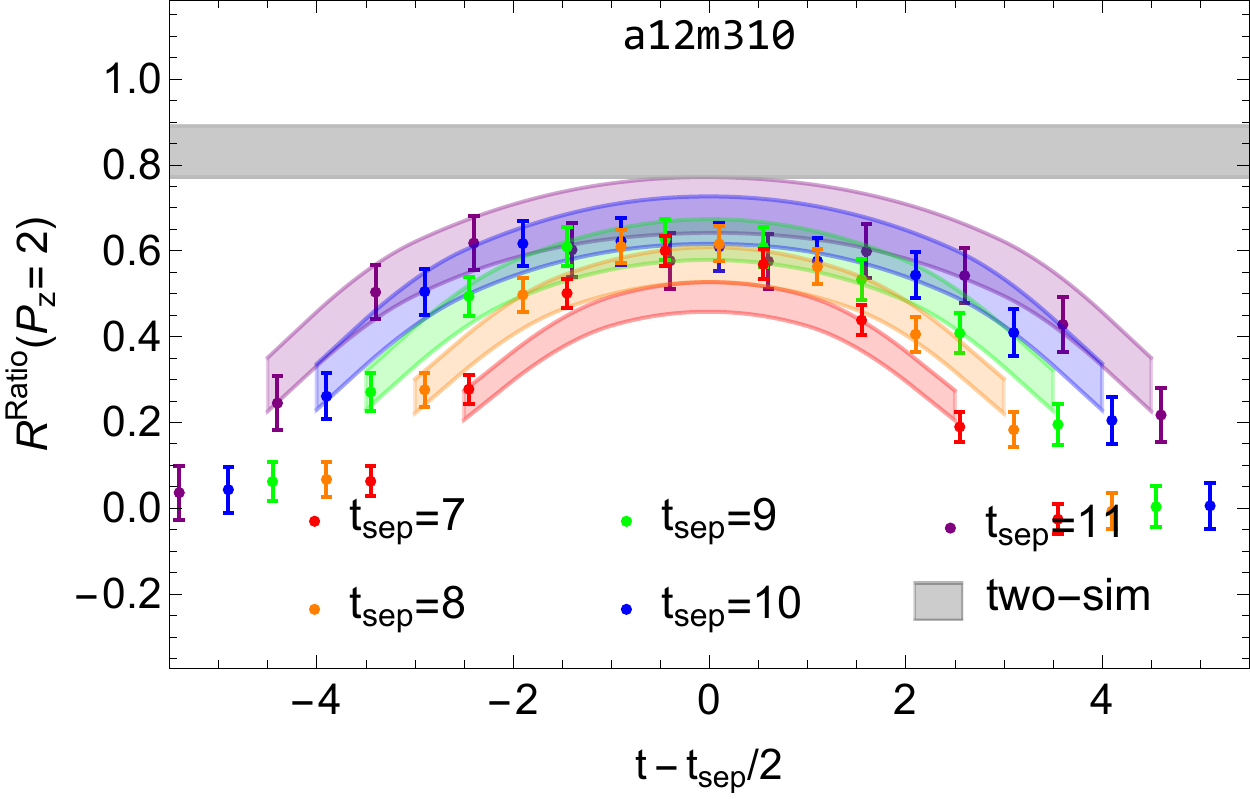}
\includegraphics[width=0.265\textwidth]{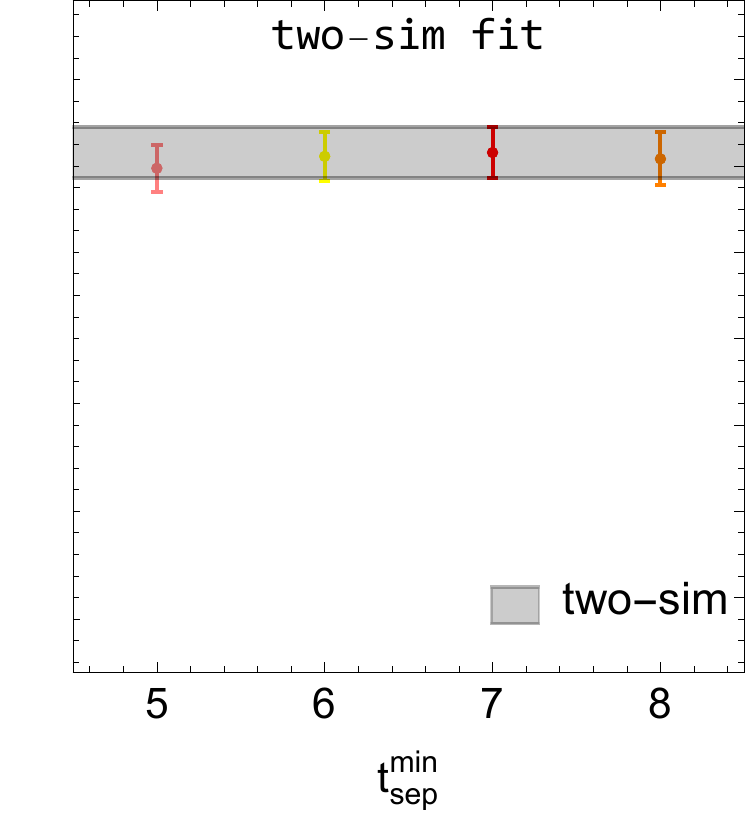}
\includegraphics[width=0.268\textwidth]{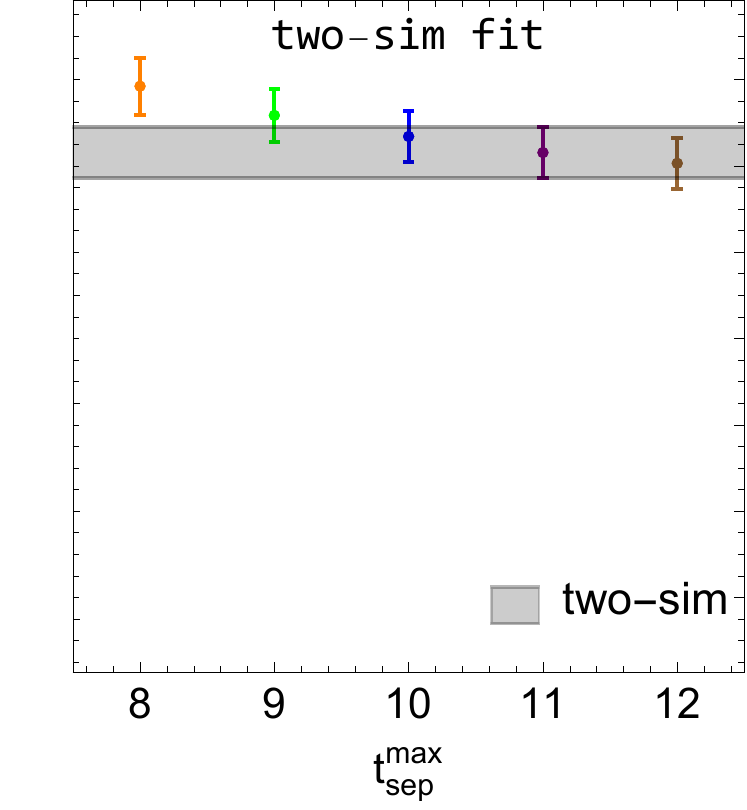}
\includegraphics[width=0.45\textwidth]{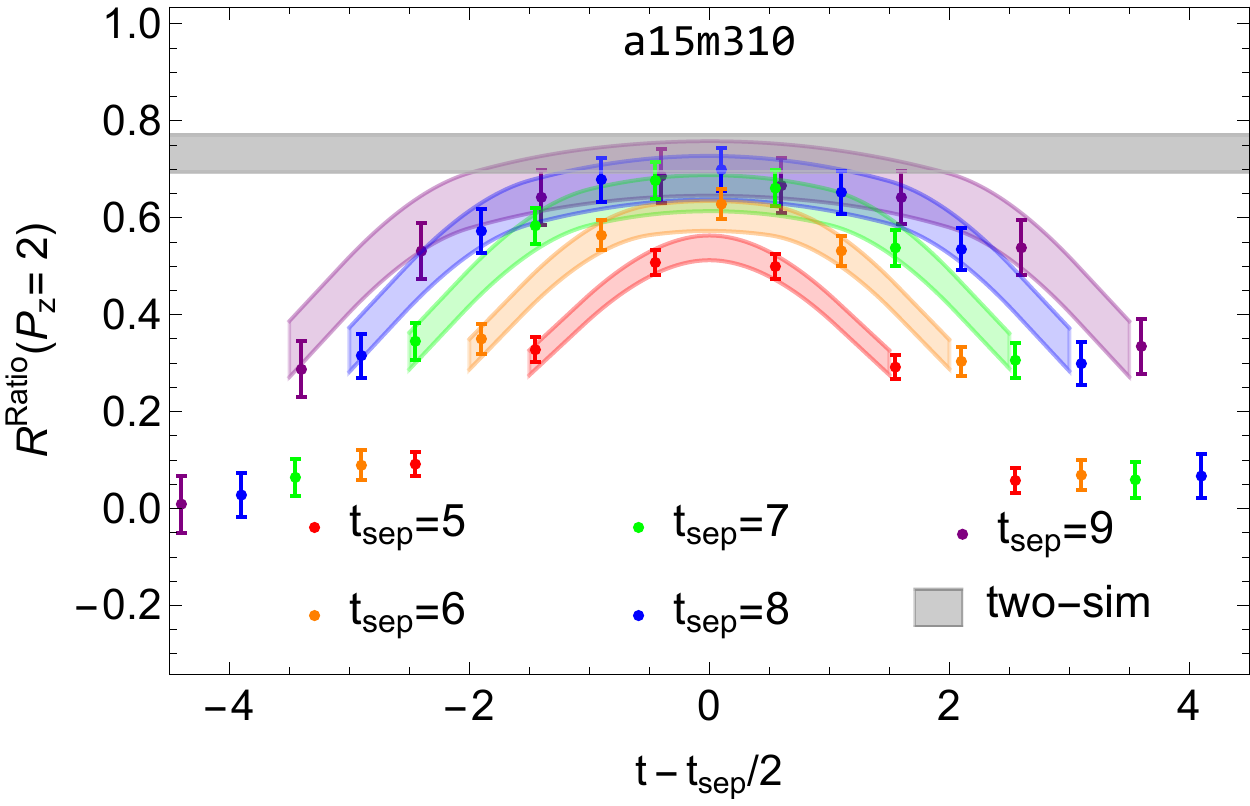}
\includegraphics[width=0.265\textwidth]{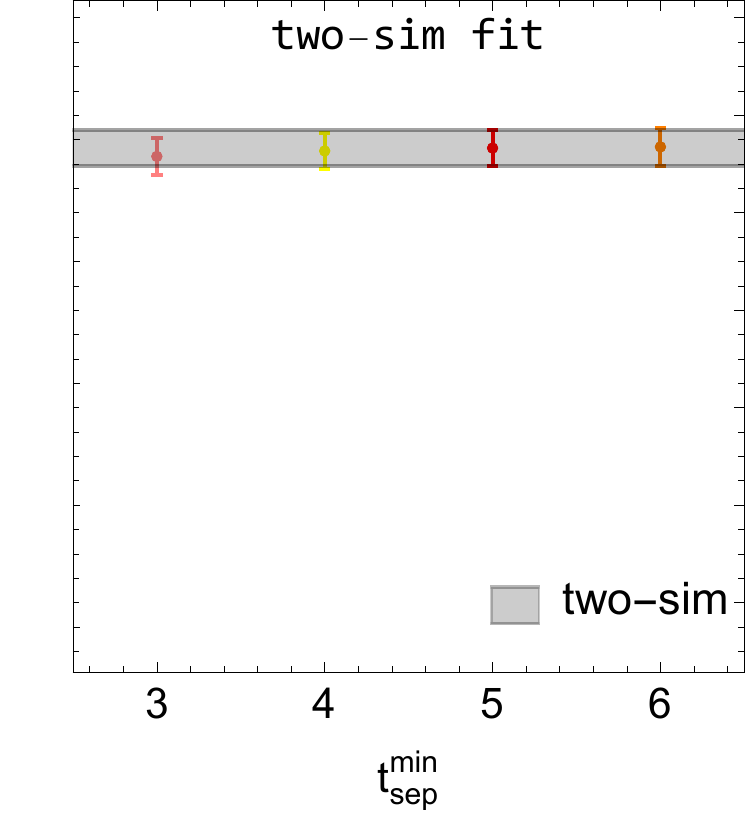}
\includegraphics[width=0.268\textwidth]{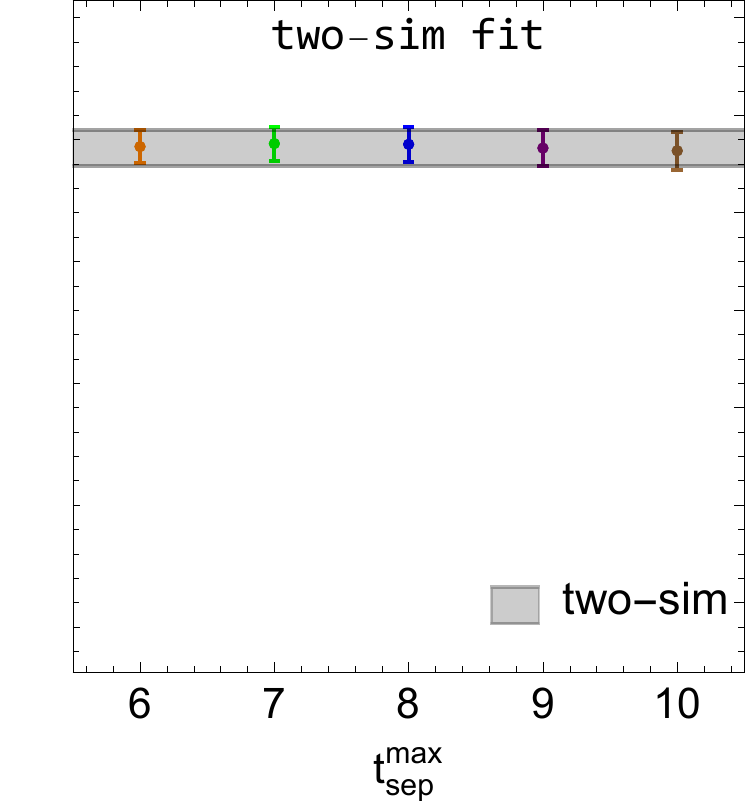}
\caption{ 
Example ratio plots (left), 
two-sim fits (right two columns) from the 
a09m310, a12m310, and a15m310 ensembles  (from top to bottom) with pion masses $M_\pi \approx 
690$~MeV respectively.
The gray bands show the extracted ground-state matrix element $\langle 0|{\cal O}|0\rangle$ obtained from a two-sim fit using $t_\text{sep}\in 
[8,12]$, $[7,11]$ and $[5,9]$ for the  
a09m310, a12m310, and a15m310 ensembles, respectively.
The first column shows the ratio of the three-point to two-point correlators with the reconstructed fit bands from the two-sim fit, shown as functions of $t-t_\text{sep}/2$. 
The seceond (third) column shows the two-sited ground-state matrix element $\langle 0|{\cal O}|0\rangle$ results with fixed $t_\text{sep}^\text{max}$ ($t_\text{sep}^\text{min}$) inputs as shown in Table~\ref{table-data} while varying $t_\text{sep}^\text{min}$ ($t_\text{sep}^\text{max}$) to see how stable the ground-state matrix elements are.
}
\label{fig:Ratio-xg-strange}
\end{figure*}

\begin{figure*}[htbp]
\centering
\includegraphics[width=0.45\textwidth]{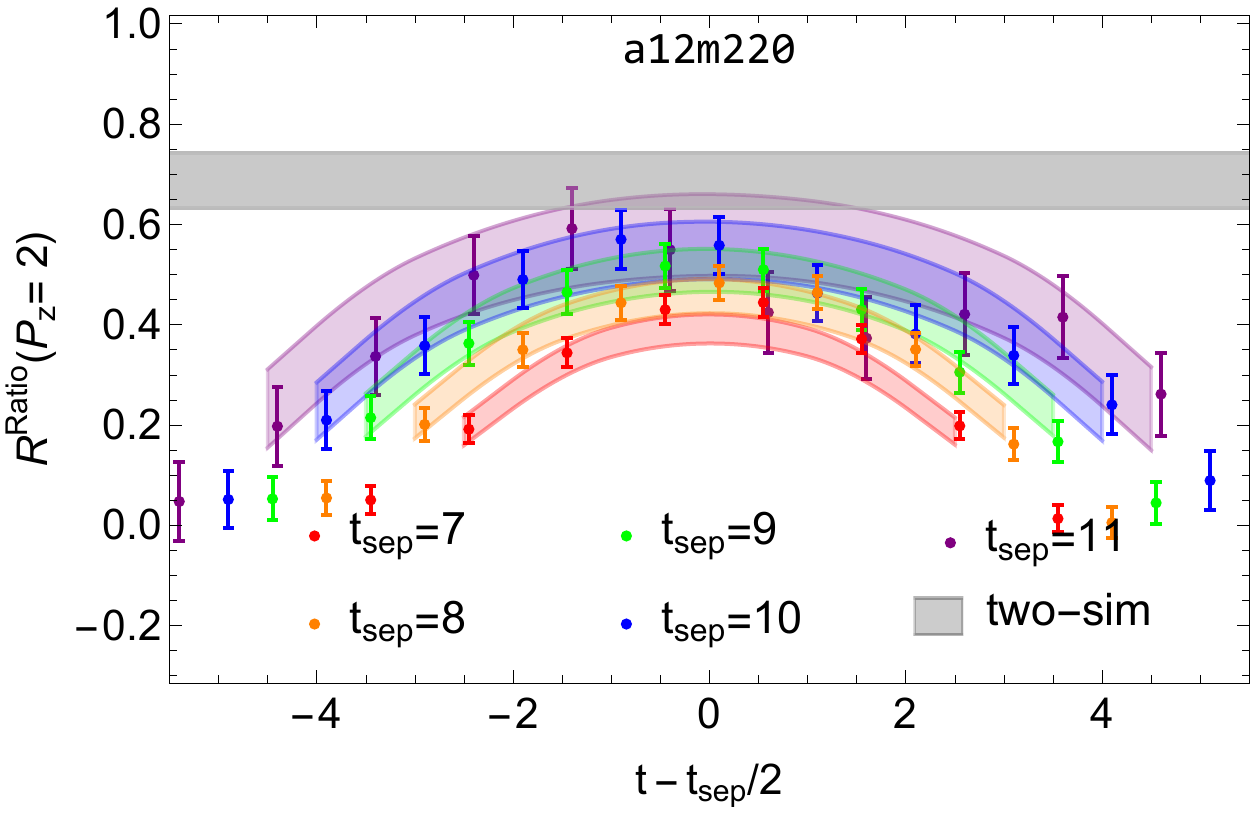}
\centering
\includegraphics[width=0.265\textwidth]{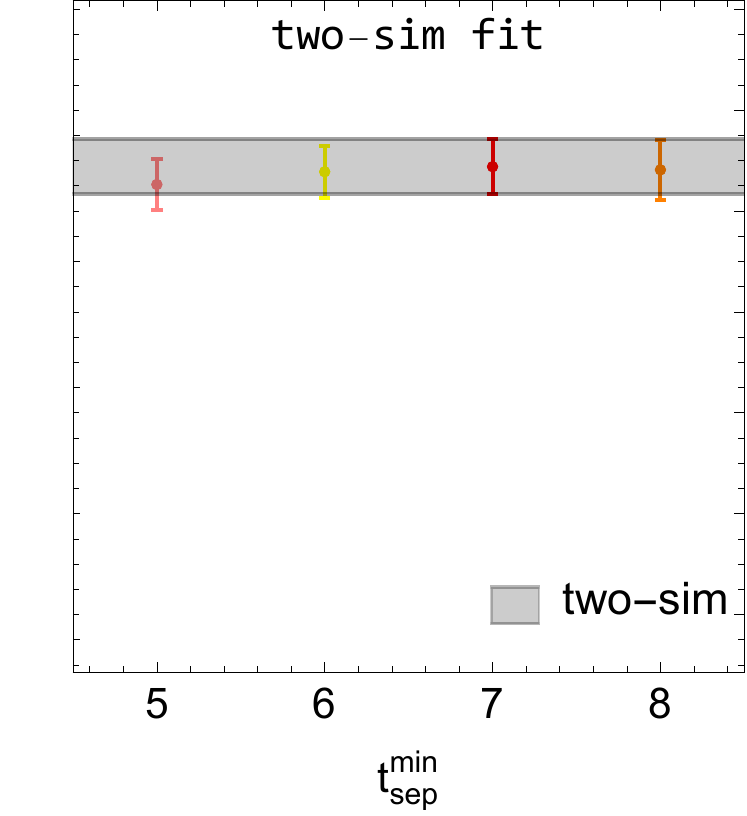}
\includegraphics[width=0.268\textwidth]{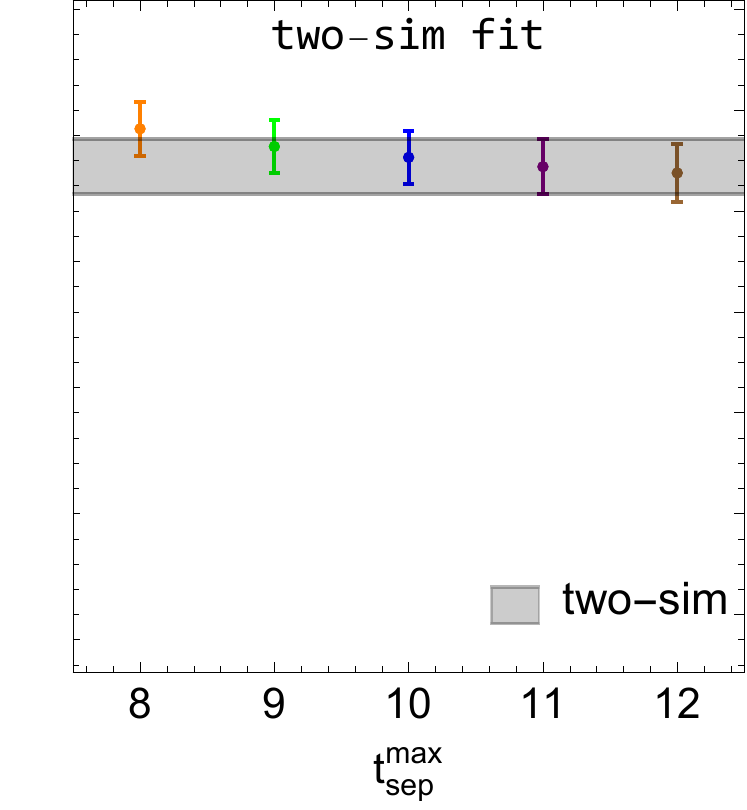}
\includegraphics[width=0.45\textwidth]{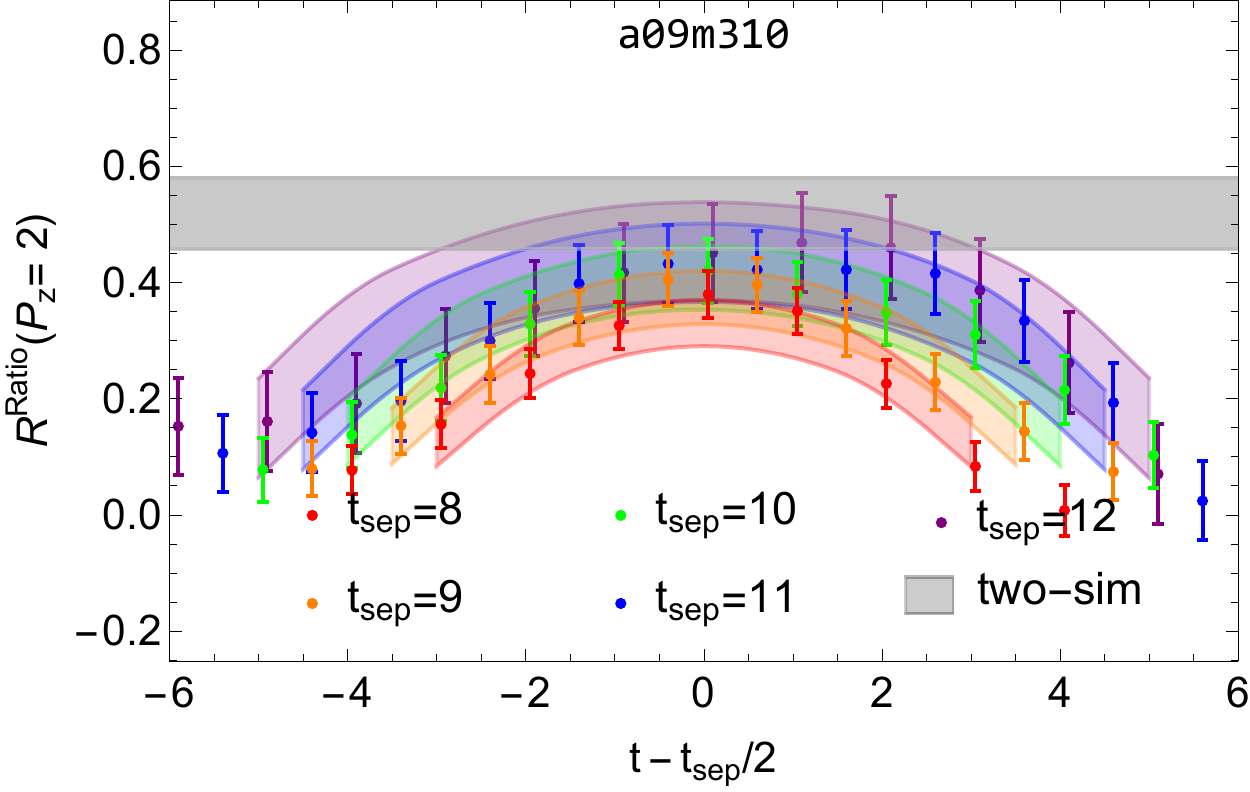}
\centering
\includegraphics[width=0.265\textwidth]{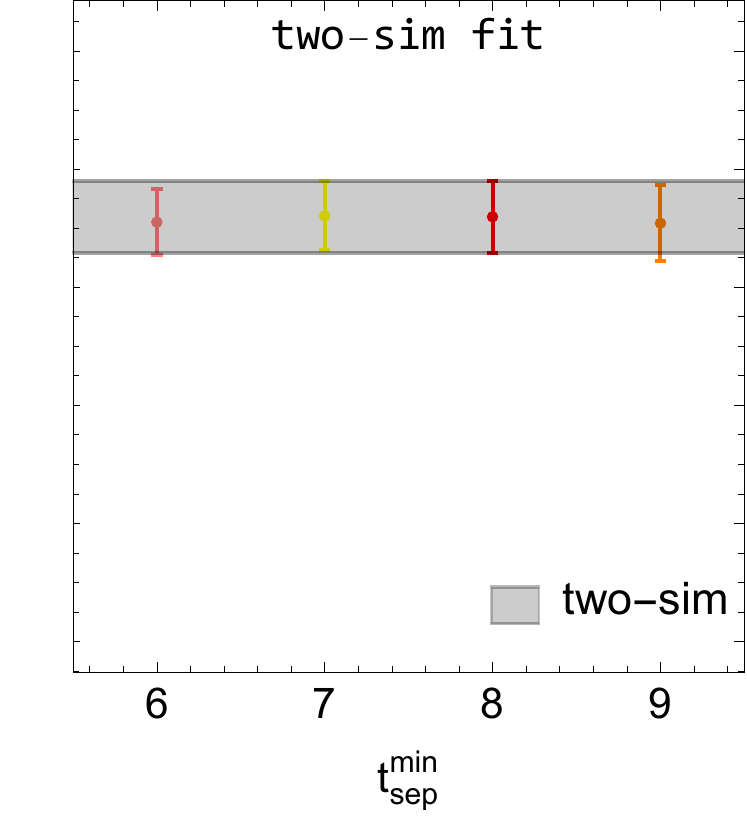}
\includegraphics[width=0.268\textwidth]{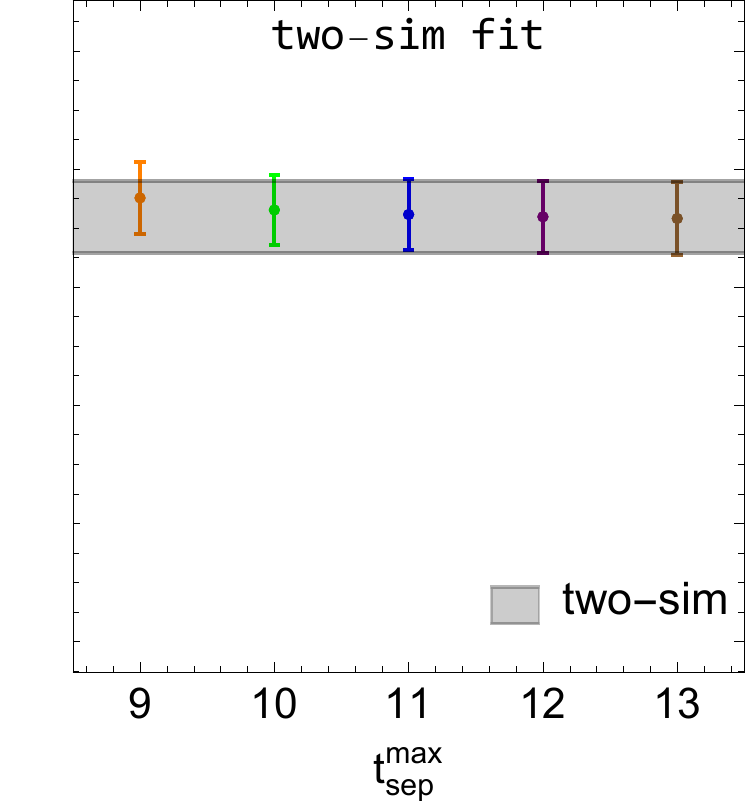}
\includegraphics[width=0.45\textwidth]{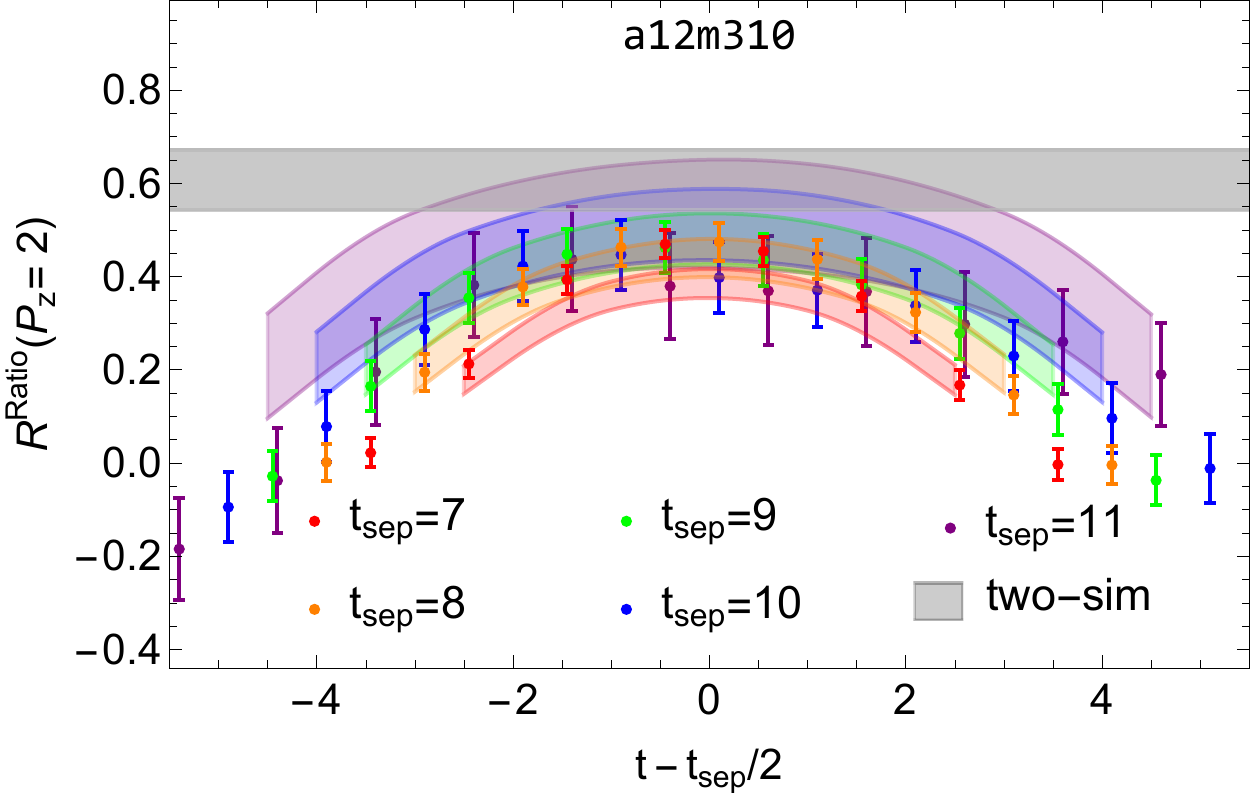}
\includegraphics[width=0.265\textwidth]{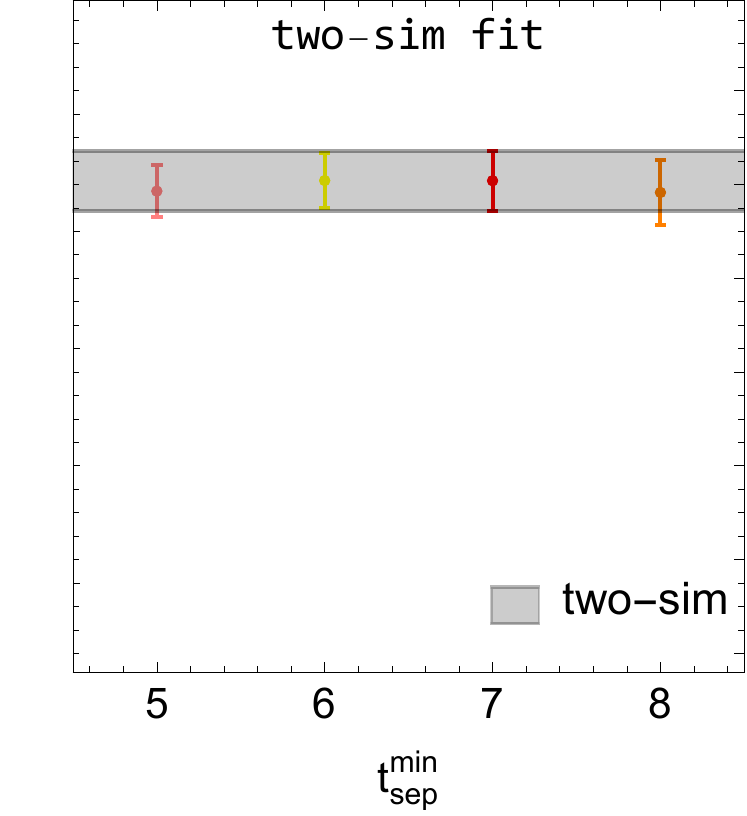}
\includegraphics[width=0.268\textwidth]{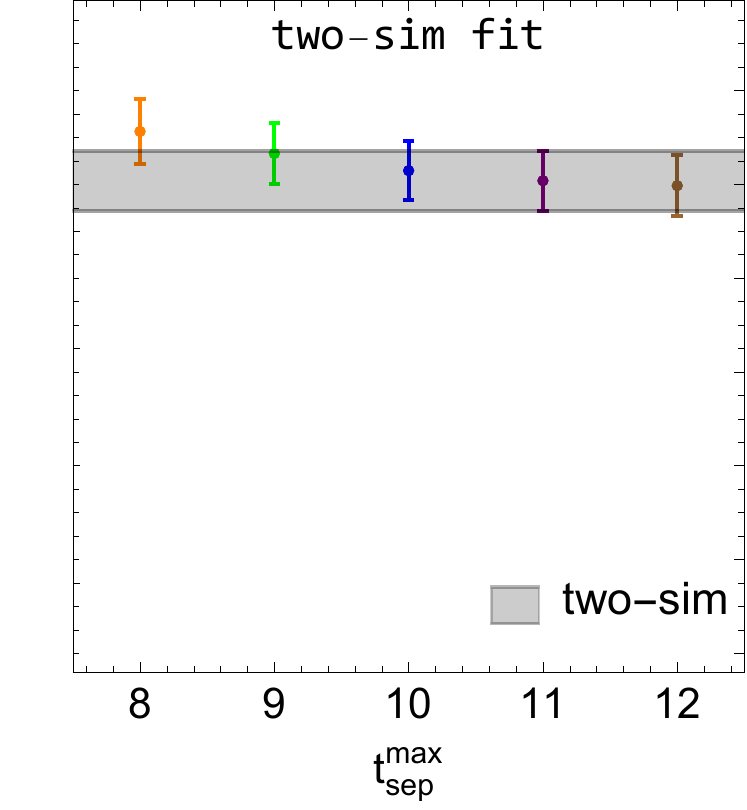}
\includegraphics[width=0.45\textwidth]{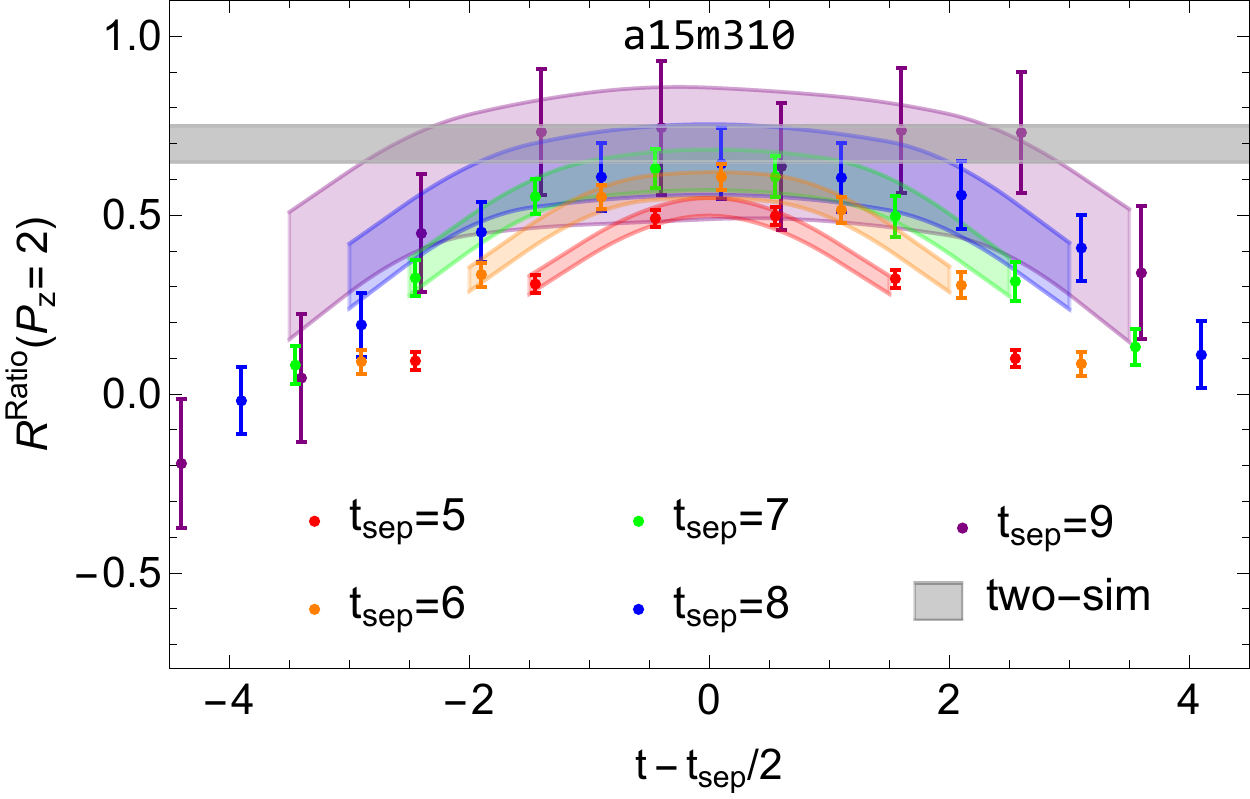}
\includegraphics[width=0.265\textwidth]{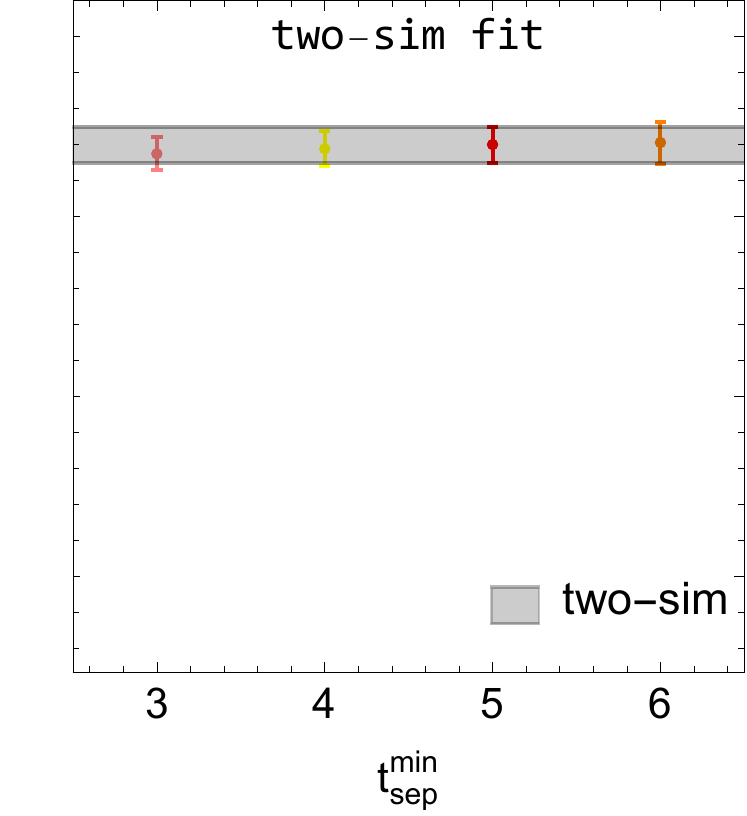}
\includegraphics[width=0.268\textwidth]{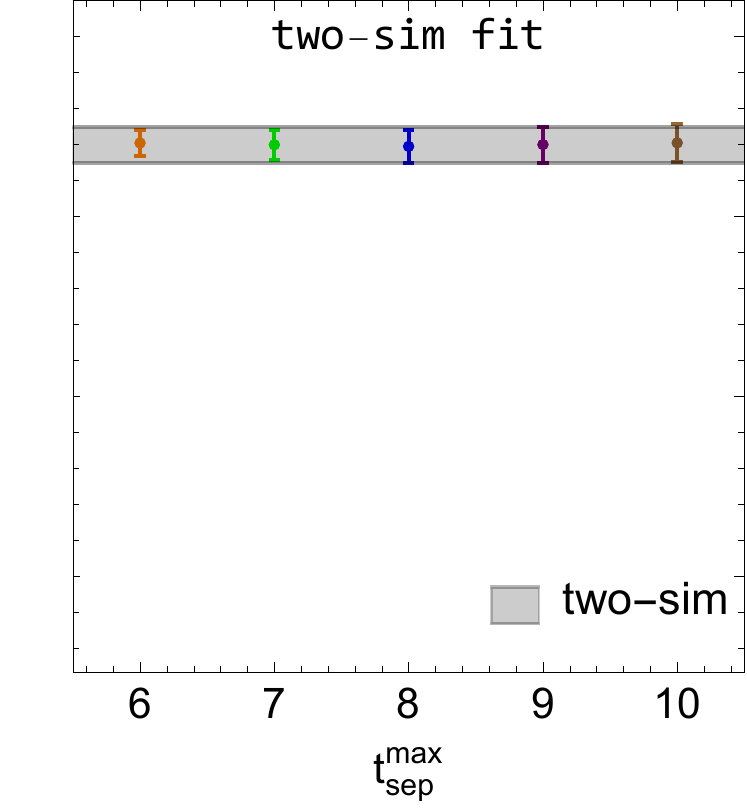}
\caption{ 
Example ratio plots (left-most column), 
and two-sim fits (right 2 columns) from the a12m220, a09m310, a12m310, a15m310 ensembles  (from top to bottom) with pion masses $M_\pi \approx \{220,310,310,310\}$ MeV respectively.
The gray bands show the extracted ground-state matrix element $\langle 0|{\cal O}|0\rangle$ obtained from
from the two-sim fit using the $t_\text{sep}\in[7,11]$, $[8,12]$, $[7,11]$, and $[5,9]$ for the a12m220, a09m310, a12m310, and a15m310 ensembles, respectively.
The first column shows the ratio of the three-point to two-point correlators with the reconstructed fit bands from the two-sim fit, shown as functions of $t-t_\text{sep}/2$. 
The second (third) column shows the two-sited  ground-state matrix element $\langle 0|{\cal O}|0\rangle$ results with fixed $t_\text{sep}^\text{max}$ ($t_\text{sep}^\text{min}$) inputs as shown in Table~\ref{table-data} while varying $t_\text{sep}^\text{min}$ ($t_\text{sep}^\text{max}$) to see how stable the ground-state matrix elements are.
}
\label{fig:Ratio-xg-light}
\end{figure*}

To visualize the quality of our fitted matrix-element extraction, we use ratios composed of the three-point ($C_N^\text{3pt}$) to the two-point ($C_N^\text{2pt}$) correlator, $R^\text{Ratio}$, defined as
\begin{equation}
\label{eq:3ptR}
R^{\text{Ratio}}_N(P_z;t_\text{sep},t)=\frac{C_N^\text{3pt}(P_z;t_\text{sep},t)}{C_N^\text{2pt}(P_z;t_\text{sep})};
\end{equation}
if the excited-state contamination were small, we would see the midpoints of $t -t_\text{sep}/2$ approach the true ground state, and these values would be independent of the $t_\text{sep}$. 
Figures~\ref{fig:Ratio-xg-strange} and~\ref{fig:Ratio-xg-light} show the bare matrix element extracted at $P_z=2$ lattice units ($2\pi P_z/(aL)$ in physical units) from three-point and two-point correlators of strange- and light-quark nucleon, respectively, for all four ensembles studied in this paper. 
The leftmost column of the figures shows the fitted ground-state gluon matrix elements  $\langle 0|{\cal O}|0\rangle$ (grey band) with multiple source-sink separation of
$R^{\text{Ratio}}$ (red to purple points) and the reconstruction of the fits to the ratio plots (red to purple bands).
We found that the $R^{\text{Ratio}}$ has a tendency to increase with larger source-sink separation $t_\text{sep}$ and toward the ground-state matrix elements obtained from the ``two-sim`` fit in Eq.~\ref{eq:3ptC} (the grey band). 
The second column of Figs.~\ref{fig:Ratio-xg-strange} and~\ref{fig:Ratio-xg-light} shows two-sim fits by fixing $t_\text{sep}^\text{max}$ at 12, 11, 11, and 9 for the a09m310, a12m220, a12m310, and a15m310 ensembles, respectively, while varying the $t_\text{sep}^\text{min}$.
We found that our ground-state matrix elements are consistent among different choices of $t_\text{sep}^\text{min}$.
Similarly, we check the dependence on $t_\text{sep}^\text{max}$ by fixing 
$t_\text{sep}^\text{min}$ of two-sim fits at 8, 7, 7, and 5 for the a09m310, a12m220, a12m310, and a15m310 ensembles respectively.
The ground-state matrix elements are mostly consistent with different choices $t_\text{sep}^\text{max}$. 
Based on the above procedure, we choose the final source-sink separation $t_\text{sep}$ (listed in Table~\ref{table-data} in lattice units) used in the ``two-sim'' fits for the rest of this work.

The majority of our two-sim fits to the three-point correlators using the parameters listed in Table~\ref{table-data} have reasonable fits with $\chi^2/\text{dof} < 1$. 
The 690-MeV a12m310 nucleon matrix elements suffer from slightly worse fits with $\chi^2/\text{dof} \approx 1.7$.
We have varied the parameters without much improvement in the quality of fit;
however, the obtained matrix elements remain consistent as long as $t_\text{sep}^\text{max}>8$.
In later sections, we will see the impact of these two matrix elements in the continuum-physical extrapolation.

\begin{figure}[htbp]
  \centering
  \includegraphics[width=0.4\textwidth]{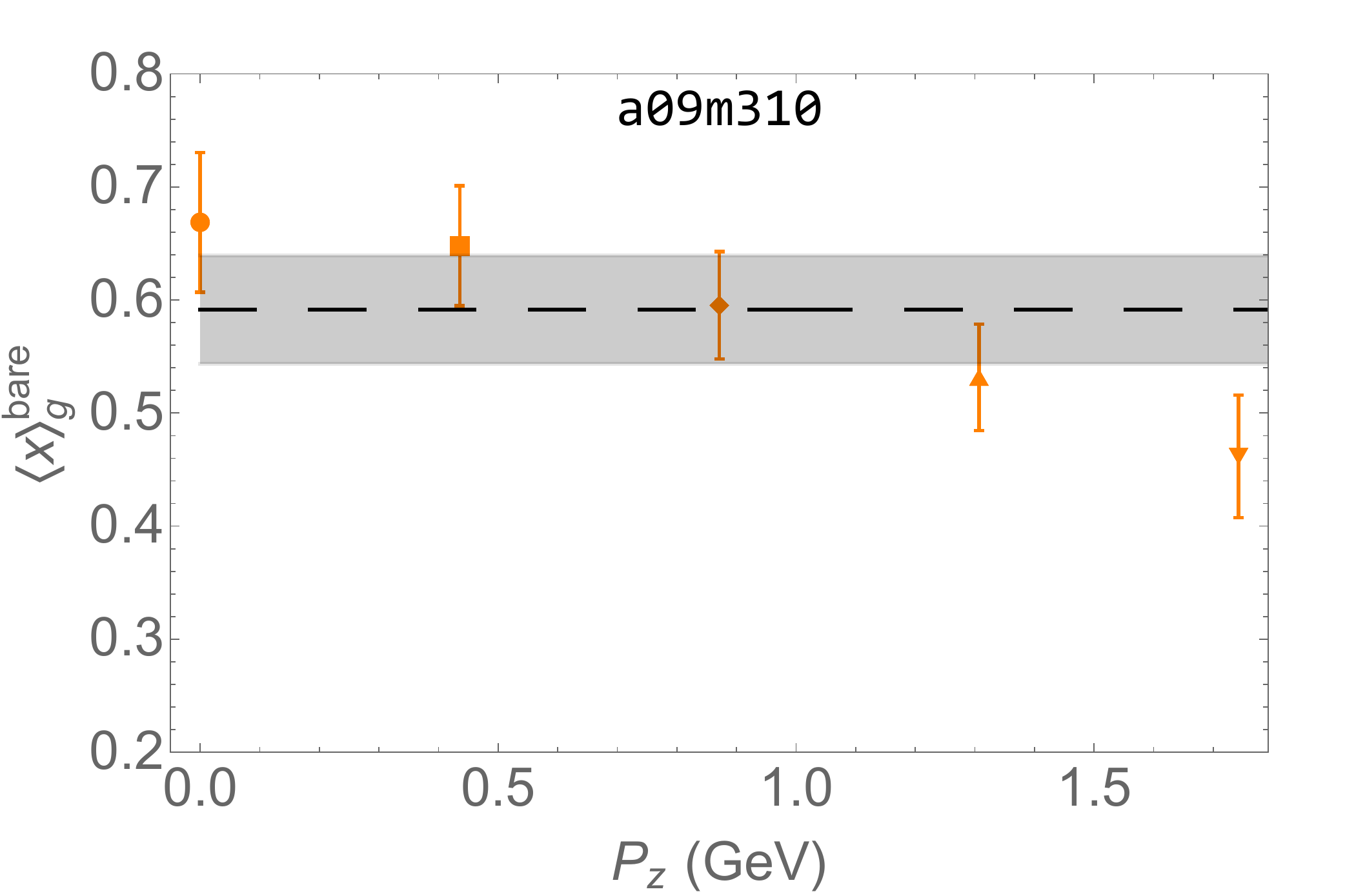}
  \includegraphics[width=0.4\textwidth]{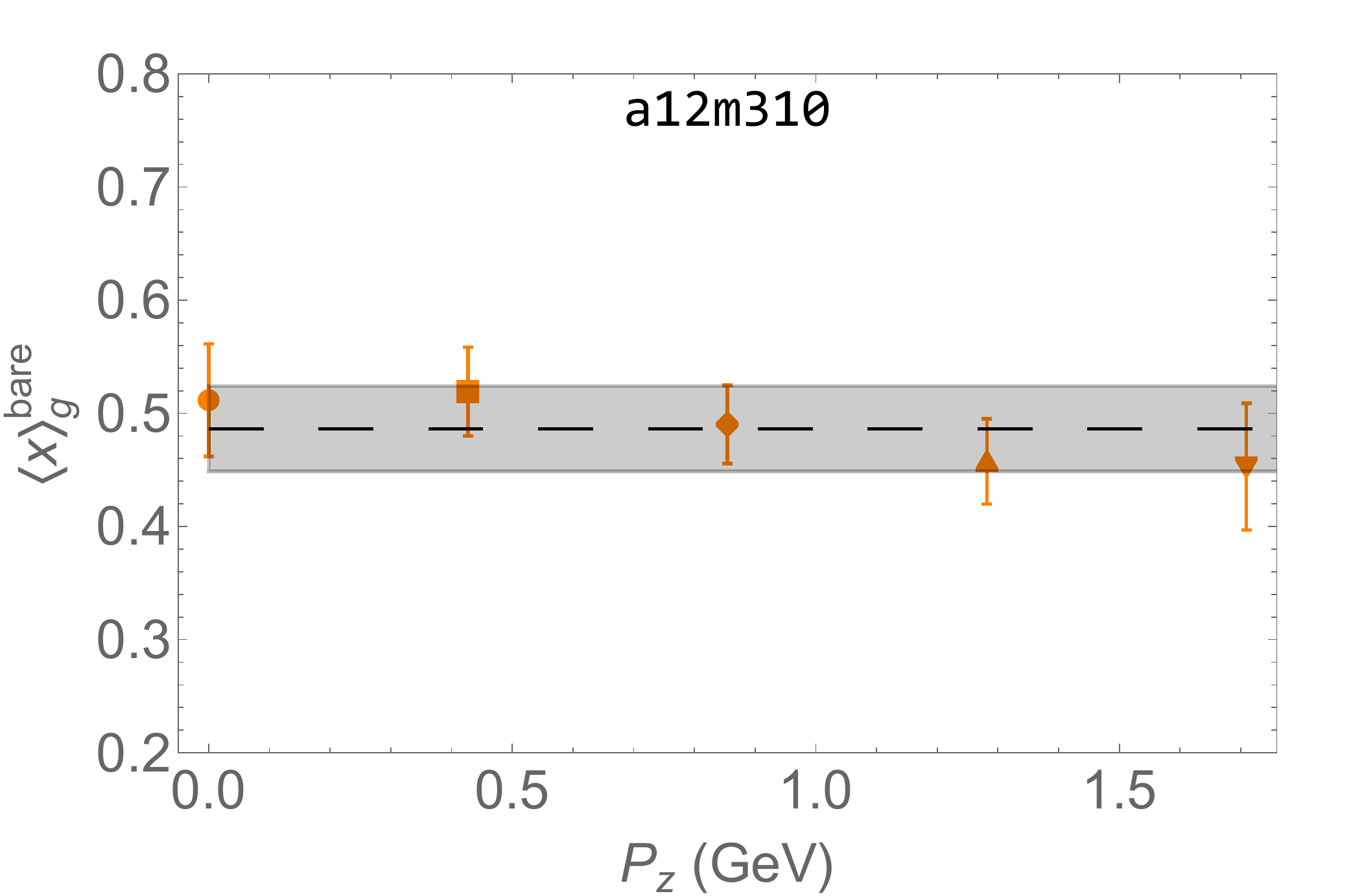}
  \includegraphics[width=0.4\textwidth]{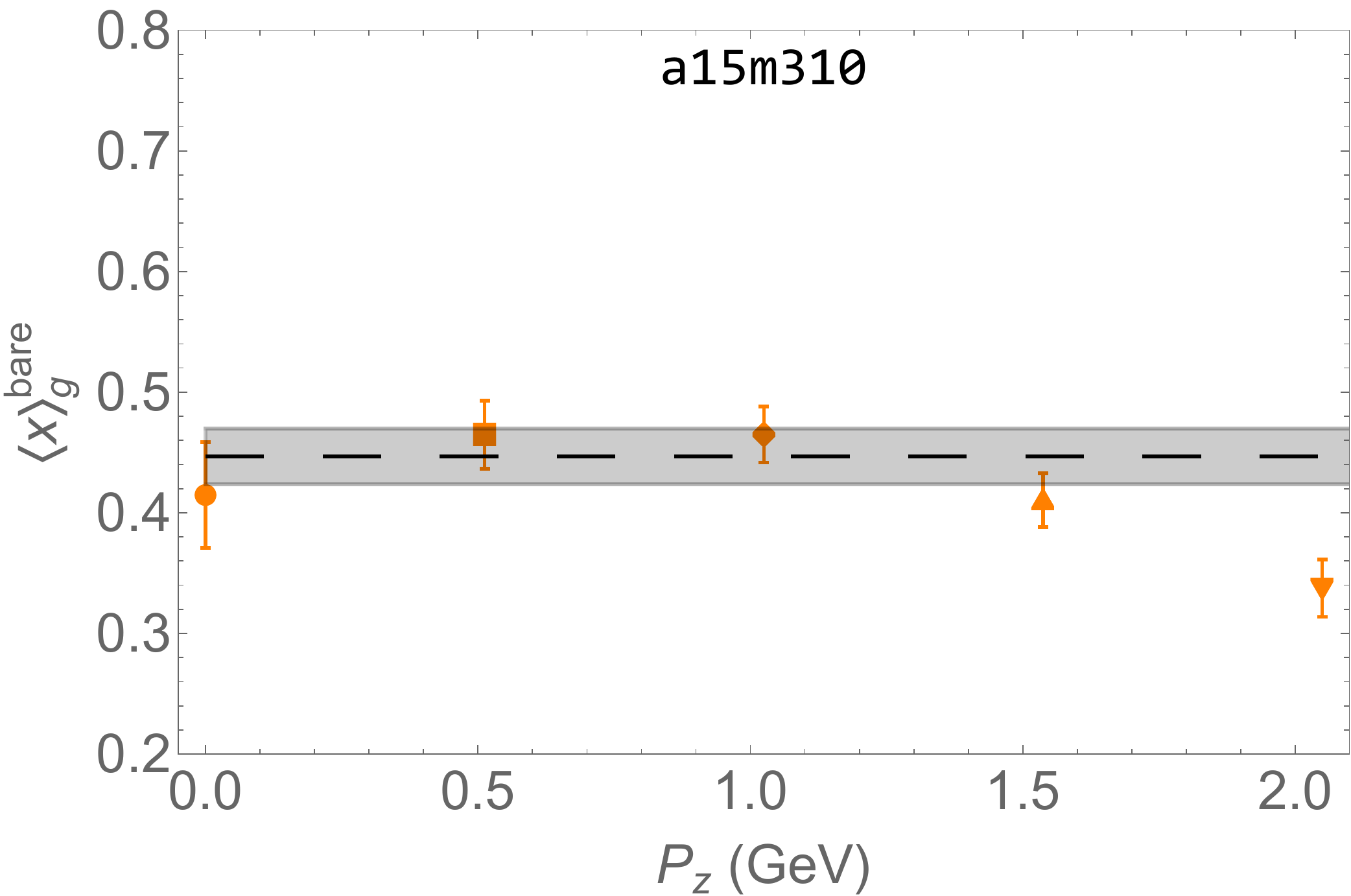}\\
  \caption{ 
  The bare gluon momentum fraction $\langle x \rangle_g^{\text{bare}}$ and fitted bands divided by kinematic factors as functions of momentum $P_z=2\pi\times N_z/L$ for $M_\pi \approx 
  690$~MeV on  
  a09m310, a12m310, and a15m310 ensembles, respectively.
   }\label{fig:bareX-strange}
\end{figure}

\begin{figure}[htbp]
  \centering
  \includegraphics[width=0.4\textwidth]{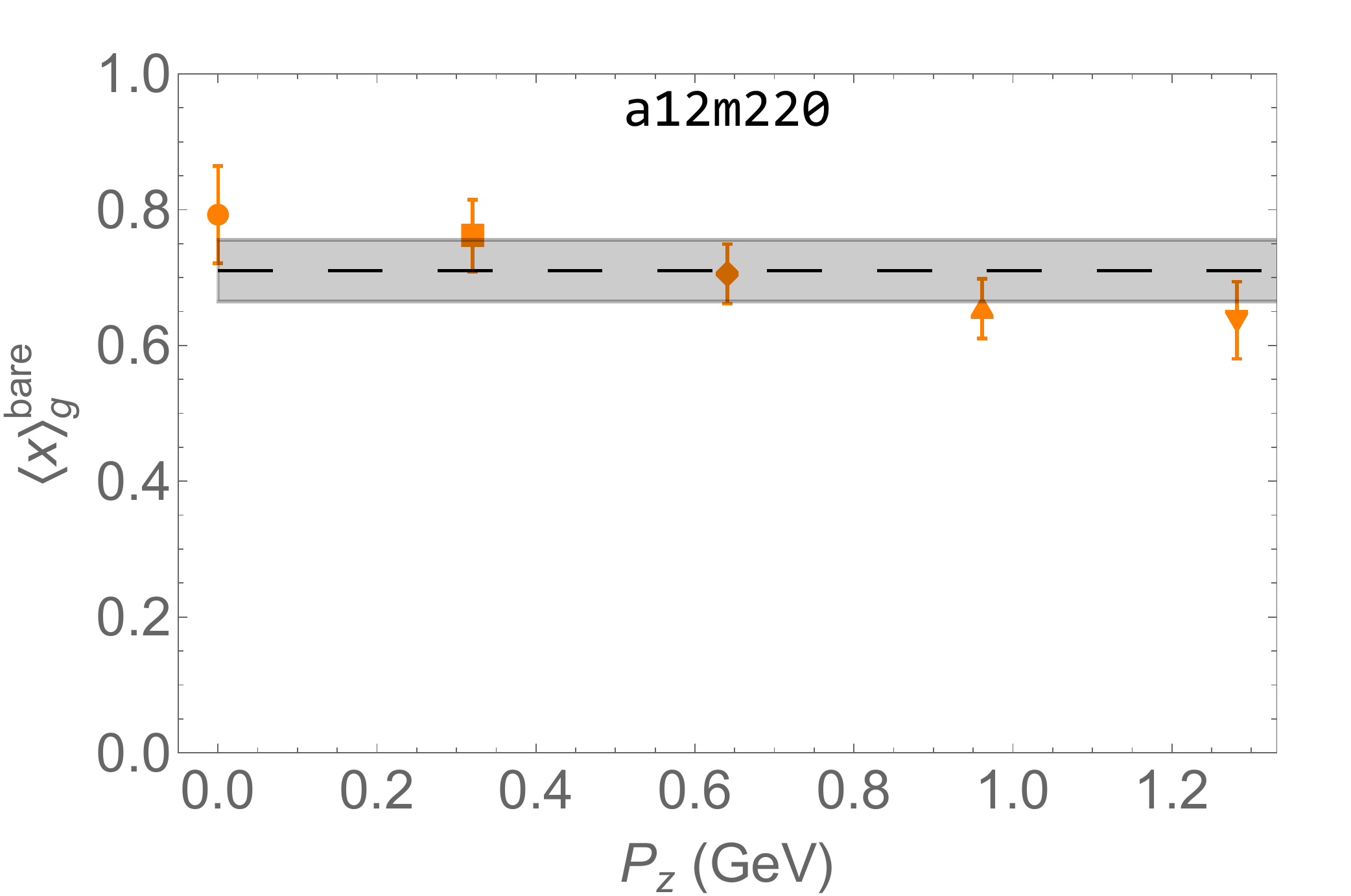}
  \includegraphics[width=0.4\textwidth]{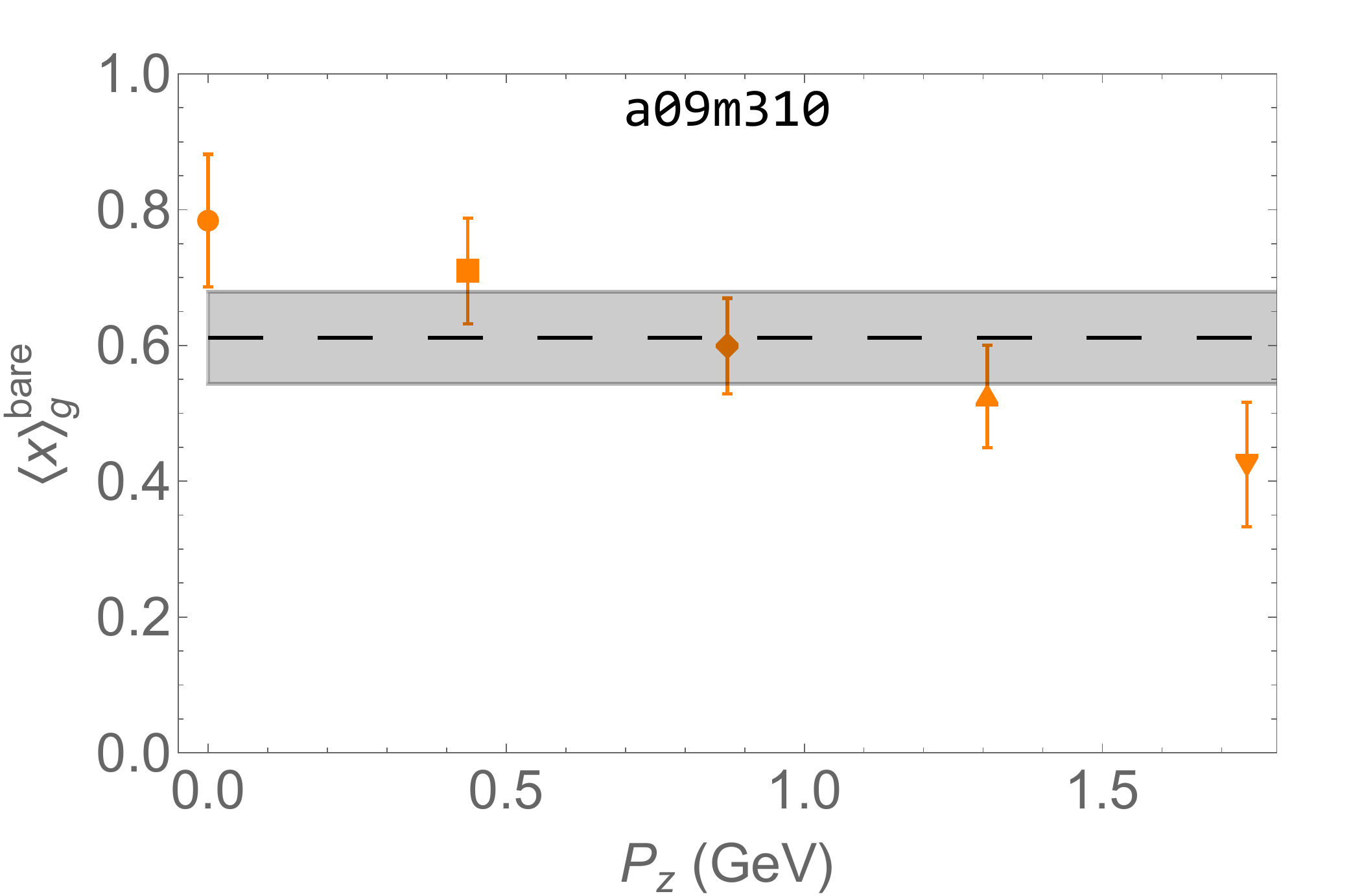}
  \includegraphics[width=0.4\textwidth]{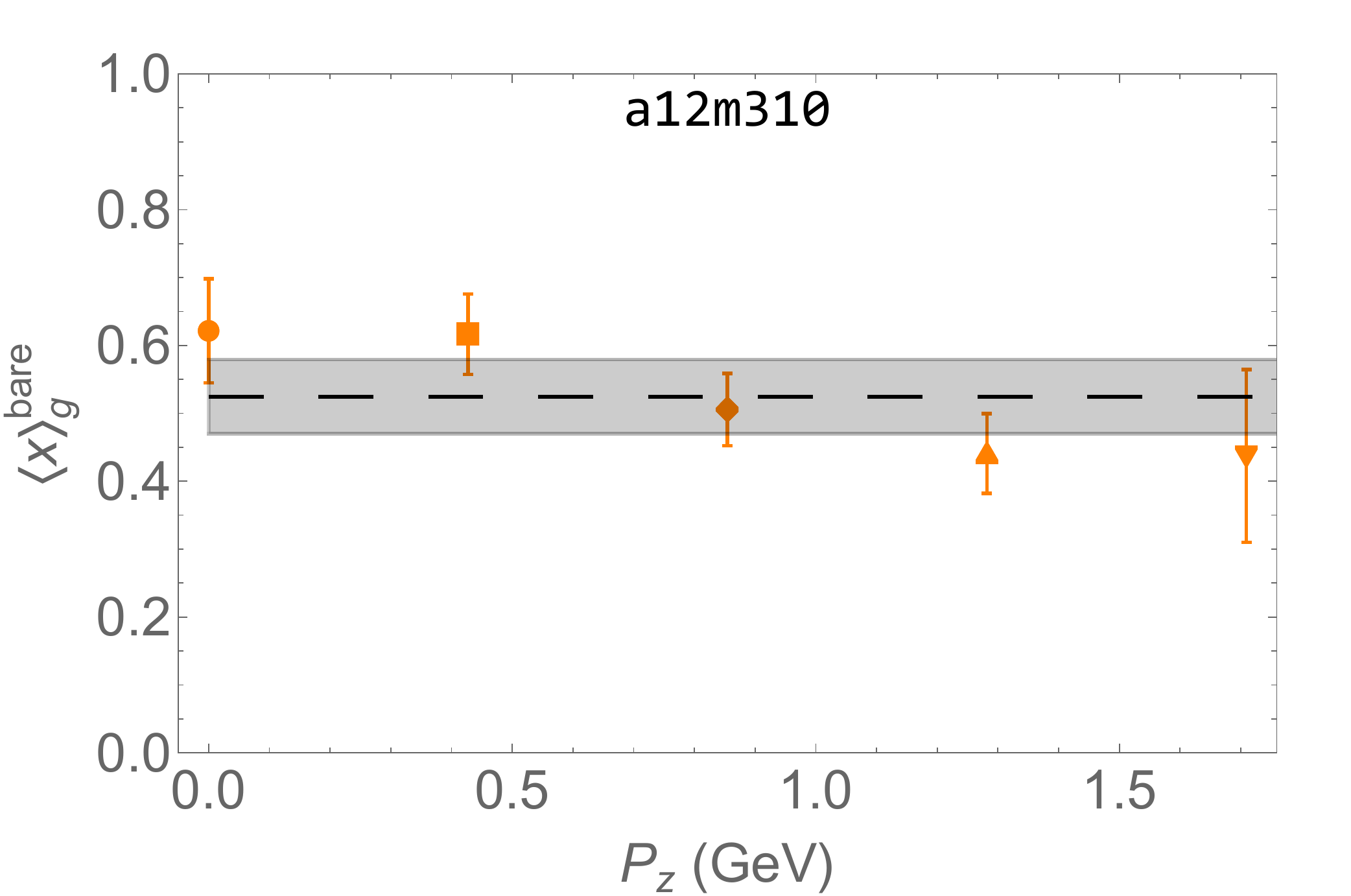}
  \includegraphics[width=0.4\textwidth]{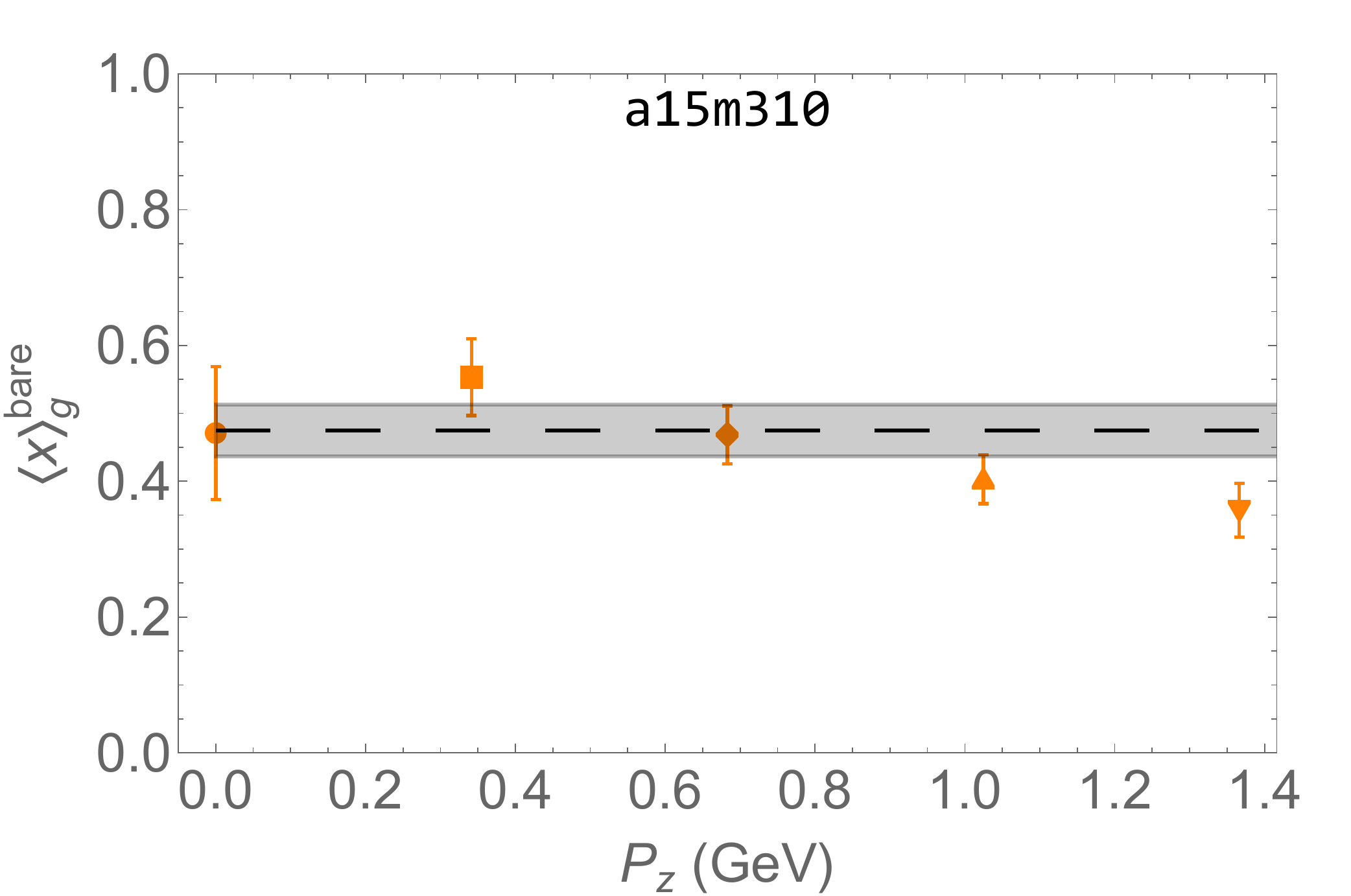}\\
  \caption{ 
  The bare gluon momentum fraction $\langle x \rangle_g^{\text{bare}}$ and fitted bands dividing by kinematic factors as functions of momentum $P_z=2\pi\times N_z/L$ for $M_\pi \approx \{220, 310, 310, 310\}$~MeV on a12m220, a09m310, a12m310, and a15m310 ensembles, respectively.
   }\label{fig:bareX-light}
\end{figure}

We repeat the same analysis routine for $P_z \in [0,4] \frac{2\pi}{L}a^{-1}$ to take advantage of the momentum-averaged results.
The above bare ground-state matrix elements $\langle 0|{\cal O}|0\rangle$ obtained from two-sim fits in Eq.~\ref{eq:3ptC} contain a kinematic factor $\frac{E_0}{\frac{3}{4}E_0^2+\frac{1}{4}P_z^2}$.
After dividing out this kinematic factor, we obtain the bare gluon momentum fraction ${\langle x \rangle_g}^{\text{bare}}$ (orange points) for four ensembles and various boost momenta, as shown in Figs.~\ref{fig:bareX-strange} and~\ref{fig:bareX-light} for strange- and light-quark nucleons.
We then fit the bare matrix elements of $P_z\in[0,4]\times 2\pi/L$ on each ensemble to a constant, shown as a gray band in the figures. 
The $\chi^2/\text{dof}$ of the fits are smaller than 1.5 except the a09m310 light nucleon fit, which is the noisiest data set and has $\chi^2/\text{dof}\approx 1.7$.
The final bare gluon momentum fractions are listed in Table~\ref{table-moments}.

\begin{table}[!htbp]
\centering
\begin{tabular}{|c|c|c|c|c|}
\hline
  ensemble &  $M_{\pi}^\text{val}$ (MeV) &  ${\langle x \rangle_g}^{\text{bare}}$ &  $\left(Z_{O_g}^{\overline{\text{MS}}}\right)^{-1}$  &  $\langle x \rangle_g^{\overline{\text{MS}}}$ \\
\hline
\multirow{1}*{a12m220} &  313.1(13) & 0.710(45) & 1.512(65)  & $0.470(30)(25)$ \\
\hline
\multirow{2}*{a09m310} & $226.6(3)$ & 0.622(63) & 1.336(106) & $0.466(46)(37)$\\
 \cline{2-5}
 ~ & $696.9(2)$ & 0.592(48) & 1.336(106) & $0.443(37)(35)$\\
\hline
\multirow{2}*{a12m310} & $309.0(11)$ & 0.651(53) & 1.512(65)  & $0.430(35)(19)$\\
 \cline{2-5}
 ~ & $684.1(6)$ & 0.637(41) & 1.512(65) & $0.421(27)(18)$\\
\hline
\multirow{2}*{a15m310} & $319.1(31)$ & 0.475(38) & 1.047(41) & $0.454(32)(18)$\\
 \cline{2-5}
 ~ & $687.3(13)$ & 0.447(23) & 1.047(41) & $0.426(20)(14)$\\
\hline
\end{tabular}
\caption{The renormalization constant $\left(Z_{O_g}^{\overline{\text{MS}}}\right)^{-1}$, the bare gluon momentum fraction ${\langle x \rangle_g}^{\text{bare}}$, and the renormalized gluon momentum fraction $\langle x \rangle_g^{\overline{\text{MS}}}$ for the four ensembles used in this calculation.
We use the a12m310 NPR factors for a12m220 $\langle x \rangle_g^{\overline{\text{MS}}}$ calculation since the mass dependence is weak for the NPR factors.
In the final column, the first error is the statistical error from the matrix element and the second error is due to the NPR factor.
}
\label{table-moments}
\end{table}

%%%%%%%%%%%%%%%%%%%%%%%%%%%%%%%%%%%%%%%%%%%%%%%%%%%%%%%%%%%%%%%%%%%%%%%%%%%%%%%%
\section{Nonperturbatively Renormalized Gluon Momentum Fraction}
\label{sec:NPR}
%%%%%%%%%%%%%%%%%%%%%%%%%%%%%%%%%%%%%%%%%%%%%%%%%%%%%%%%%%%%%%%%%%%%%%%%%%%%%%%%

\begin{figure}[htbp]
  \centering
  \includegraphics[width=0.46\textwidth]{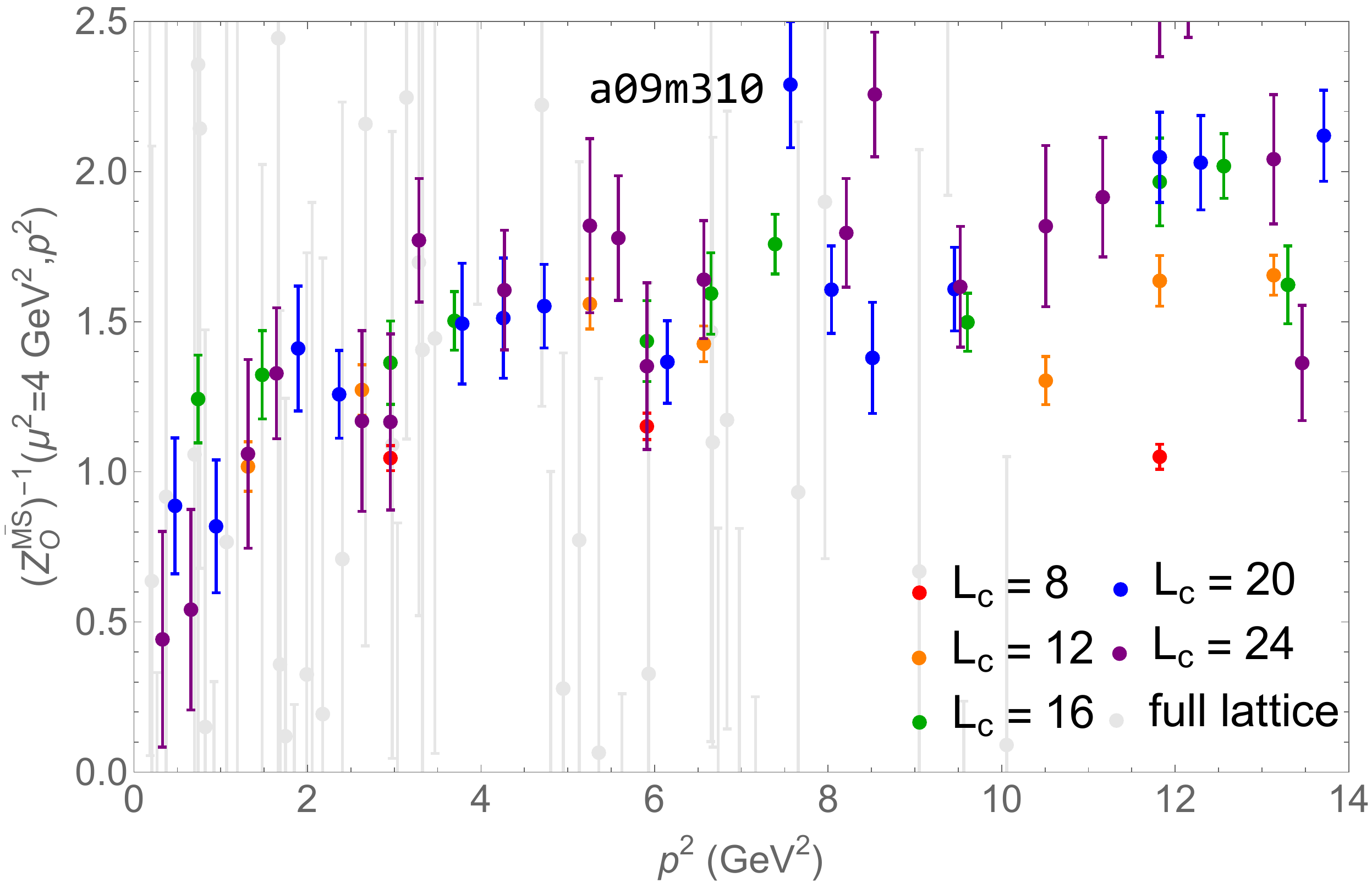}
  \includegraphics[width=0.46\textwidth]{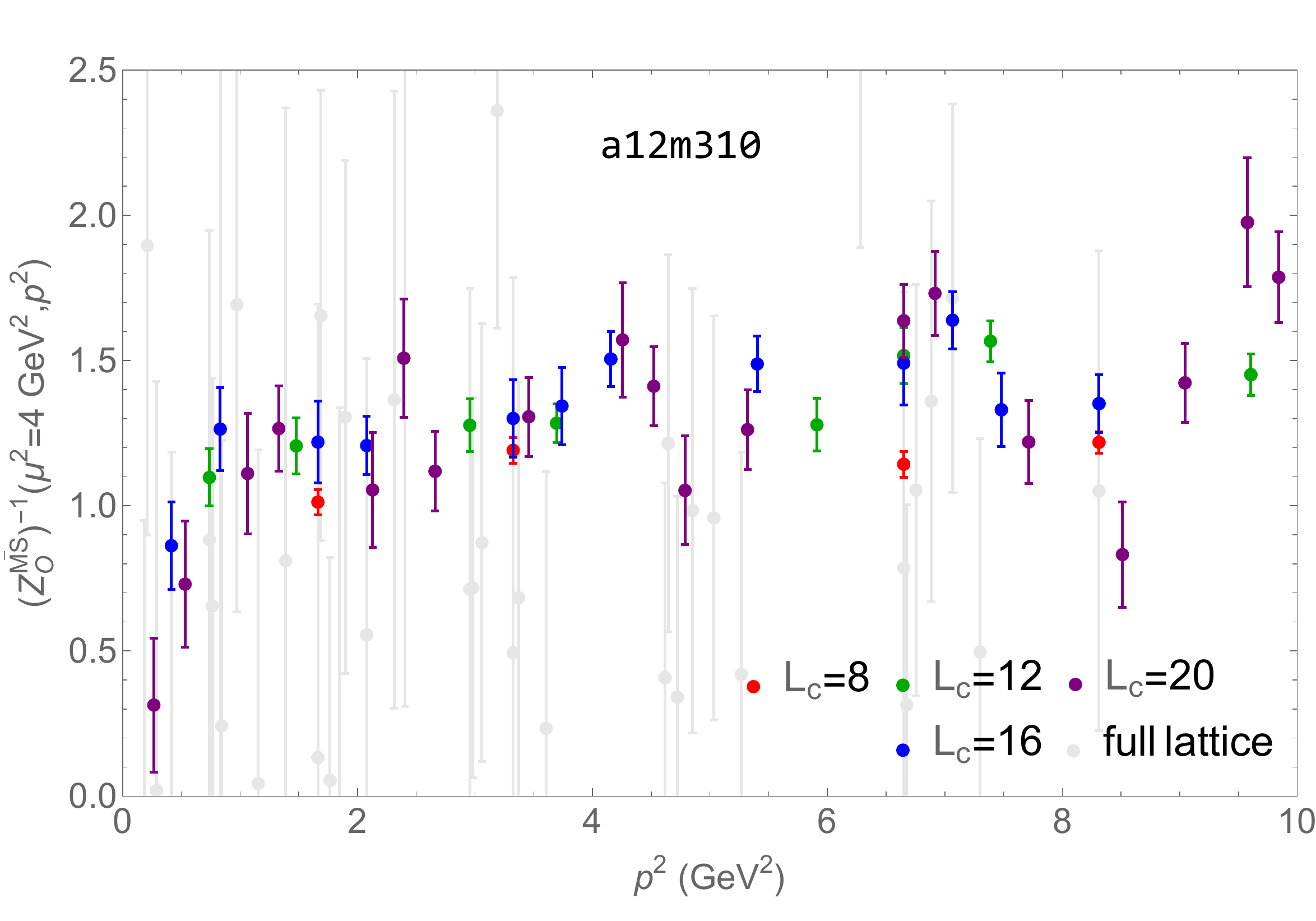}
  \includegraphics[width=0.46\textwidth]{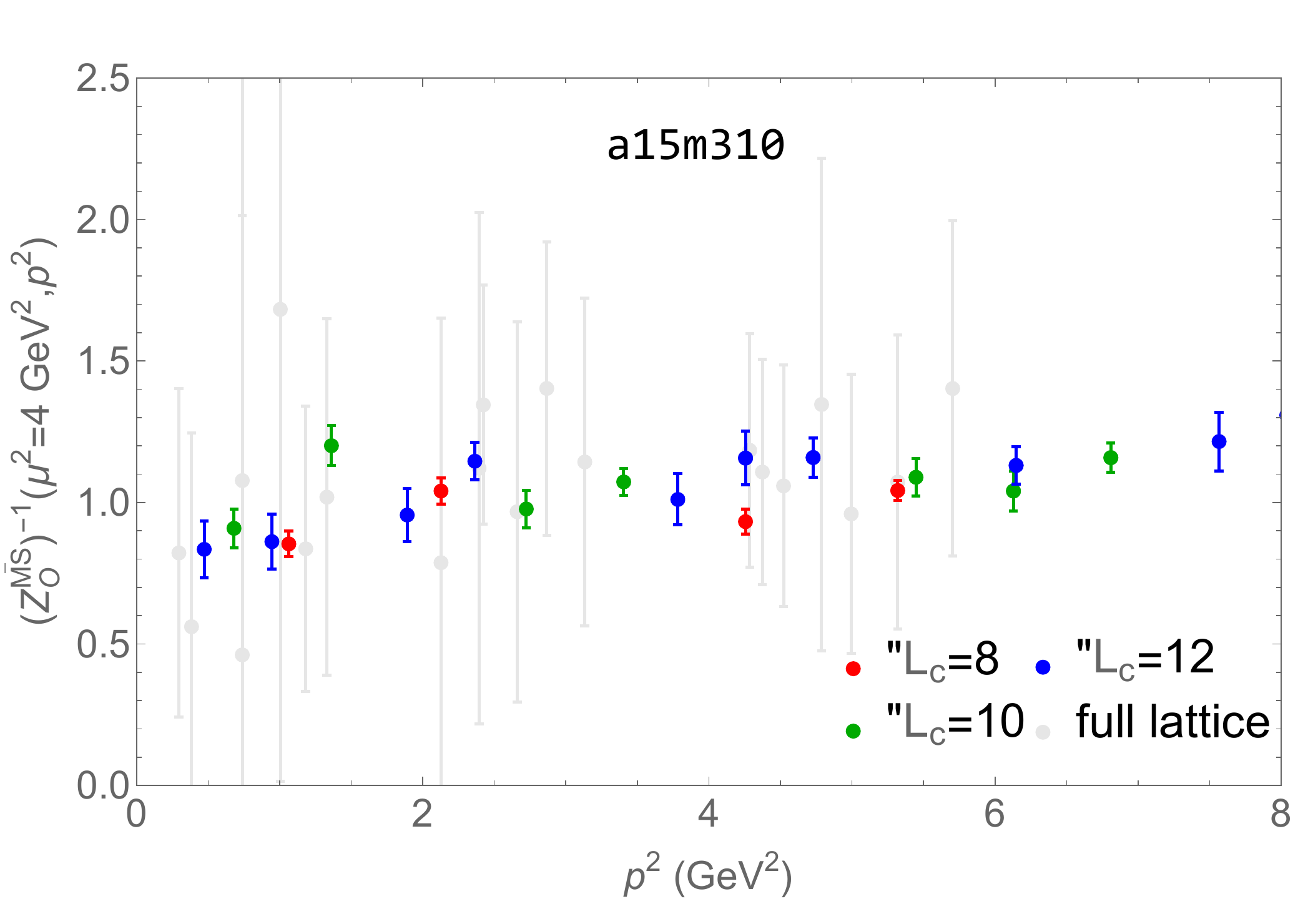} 
  \caption{ 
  The renormalization constants $\left(Z_{O_g}^{\overline{\text{MS}}}\right)^{-1}(\mu=4\text{ GeV}^2,p^2)$ as a function of $p^2 (\text{ GeV}^2) $ for the a09m310, a12m310, and a15m310 ensembles 
  are shown in the first second, and last rows, respectively. Different color points represent different cutoffs $L_c$ and the lighter gray large error-bar points are from the full lattice calculations. 
   }\label{fig:Z-L-dep}
\end{figure}

After we determine the gluon bare momentum fraction matrix element  from lattice calculation, our next step is to renormalize it.
In this work, we will be using RI-MOM--scheme NPR~\cite{Martinelli:1994ty}.
We then implement a perturbative matching to convert the gluon momentum fraction into the $\overline{\text{MS}}$ scheme as follows
\begin{align}
\langle x \rangle_g^{\overline{\text{MS}}} &= Z_{O_g}^{\overline{\text{MS}}}(\mu^2,\mu^2_R){\langle x \rangle_g}^{\text{bare}} \nonumber\\
&= R^{\overline{\text{MS}}}(\mu^2,\mu^2_R)Z^\text{RI}_{O_g}(\mu^2_R){\langle x \rangle_g}^{\text{bare}},
\label{eq:xgMSbar}
\end{align}
where $Z_{O_g}^{\overline{\text{MS}}}(\mu^2,\mu^2_R)$ is the renormalization constant, and the one-loop expression for the perturbative matching ratio $R^{\overline{\text{MS}}}(\mu^2,\mu^2_R)$, derived in Ref.~\cite{Yang:2016xsb}, is
\begin{align}
R^{\overline{\text{MS}}}(\mu^2,\mu^2_R)&= 1-\frac{g^2N_f}{16\pi^2}\left(\frac{2}{3}\log(\mu^2/\mu^2_R)+\frac{10}{9}\right) \nonumber \\
&- \frac{g^2N_c}{16\pi^2}\left(\frac{4}{3}-2\xi+\frac{\xi^2}{4}\right),
%\label{}
\end{align}
where the number of flavors $N_f=4$, the number of colors $N_c=3$, the parameter from the Riemann zeta function $\xi=0$ in the Landau gauge, $g^2$ is $4\pi\alpha(\mu)$~\cite{Herren:2017osy,Schmidt:2012az,Chetyrkin:2000yt}, and $\mu=2$~GeV is used in our calculation.
The RI-MOM renormalization factor $Z^\text{RI}_{O_g}(\mu^2_R)$ can be obtained with the condition
\begin{equation}
Z_g(p^2)Z^\text{RI}_{O_g}(p^2)\Lambda^{\text{bare}}_{O_g}(p)(\Lambda^{\text{tree}}_{O_g}(p))^{-1}|_{p^2=\mu^2_R}=1,
\end{equation}
where $Z_g(p^2)$ is the gluon-field renormalization and $\Lambda^{\text{bare (tree)}}_{O_g}$ is the bare (tree-level) amputated Green function for the operator $O_g$ in the Landau-gauge--fixed gluon state. 
The NPR factor $Z^\text{RI}_{O_g}(p^2)$ of the operator in Eq.~\ref{eq:op_def} 
is derived in Ref.~\cite{Shanahan:2018pib,Yang:2018bft}:
\begin{align}\label{eq:ZRI}
&(Z^\text{RI}_{O_g})^{-1}(\mu_R^2)\\ \nonumber
&=\left.\frac{p^2\langle (O_{g,\mu\mu}-O_{g,\nu\nu})\Tr[A_\tau(p)A_\tau(-p)] \rangle}{2(p_\mu^2-p_\nu^2)D_{g,\tau\tau}(p)}\right|_{p^2=\mu_R^2, \tau\neq\mu\neq\nu, p_\tau=0}.
\end{align}
Therefore, the gluon propagator $D_{g,\mu\nu}(p)$ and bare gluon amputated Green function $\Lambda_{O_g}^\text{bare}(p)$ need to be calculated for the further calculation of the NPR factor:
\begin{align}
D_{g,\mu\nu}(p) &=\langle \Tr[A_\mu(p)A_\nu(-p)] \rangle \nonumber \\
\Lambda_{O_g}^\text{bare}(p)
&=\frac{\langle (O_{g,\mu\mu}-O_{g,\nu\nu})\Tr[A_\tau(p)A_\tau(-p)] \rangle(N_c^2-1)^2}{4D_{g,\tau\tau}^2(p)},
\label{}
\end{align}
where $\tau, \mu, \nu\in\{x,y,z,t\}$ and $\tau\neq\mu\neq\nu$.
Following the above procedure, $Z_{O_g}^{\overline{\text{MS}}}(\mu^2=4 \text{ GeV}^2,p^2)$ is calculated and shown in Fig.~\ref{fig:Z-L-dep} in light gray points by using the full lattice of all ensembles listed in Table~\ref{table-data}.
The signal-to-noise ratios of the light gray points are larger than 100\% in most cases, which gives us a useless renormalized gluon momentum fraction.
The relative errors also become larger as the lattice spacing becomes smaller.
For example, the relative errors of $Z_{O_g}^{\overline{\text{MS}}}(\mu^2=4 \text{ GeV}^2,p^2)$ for a09m310 ensemble are $\approx 1.5$ on $347$ configurations.
To achieve a comparable relative error as the bare matrix elements of the light nucleon ($0.10$) shown in Table~\ref{table-moments}, we need $15^2\times 347=78,075$ configurations for the a09m310 NPR calculation alone, which is very expensive to do in dynamical gauge generation.
Therefore, we need some technique to reduce the error of the NPR factor without requiring a huge number of configurations in the calculation.

In Refs.~\cite{Liu:2017man,Yang:2018bft}, $\chi$QCD introduces a technique called cluster-decomposition error reduction (CDER) in order to increase the signal-to-noise ratio of NPR factor, which has not been widely used by other lattice groups.  
The reason for such error reduction is that, for the operator insertions, the correlator signal falls off exponentially with the distance, while the error remains constant.
Beyond a certain correlation length, it will only increase the noise without gaining any signal. 
$\chi$QCD introduced two additional cutoffs in the CDER technique~\cite{Yang:2018bft} for calculating the gluon NPR: $r_1$ ($r_2$) for the upper bound of the distance between the glue operator and one of the gauge fields (the gauge fields in the gluon propagator $D_{g}(p)$) in the gluon amputated Green function $\Lambda_{O_g}(p)$ definition.
With these two cutoffs, the correlators in the gluon propagator and gluon amputated Green function become 
\begin{align}
&\langle \Tr[A_\mu(p)A_\nu(-p)] \rangle \nonumber\\
&\approx \Big\langle \int_{|r'|<r_2}\!\!\!\!\!\!d^4r' \int d^4x\,
e^{ip\cdot r'} \Tr[A_{\mu}(x) A_{\nu}(x+r')] \Big\rangle,
\end{align}
\begin{align}
&\langle (O_{g,\mu\mu}-O_{g,\nu\nu})\Tr[A_\tau(p)A_\tau(-p)] \rangle \nonumber\\
&\approx \Big\langle \int_{|r|<r_1}\!\!\!\!\!\!d^4r \int_{|r'|<r_2}\!\!\!\!\!\!d^4r' \int d^4x\,
e^{ip\cdot r'} \nonumber\\
&\times [O_{g,\mu\mu}-O_{g,\nu\nu}](x+r)\Tr[A_{\tau}(x) A_{\tau}(x+r')] \Big\rangle. \label{eq:def_cder}
\end{align} 
Reference~\cite{Yang:2018bft} studies the gluon nonperturbative renormalization on different types of gauge configurations:
2+1-flavor RBC/UKQCD domain-wall fermion (DWF) 
with lattice spacing $a=0.114$~fm, $m_\pi=140$~MeV 
, a quenched Wilson gauge ensemble of 0.098~fm,
and two volumes of 0.117~fm 450-MeV two-flavor clover fermion as well.
In their quenched and two-flavor clover fermion studies, 
they compare the NPR-factor $Z^\text{RI}_{O_g}$ results using the CDER technique and $100\times$ statistics and show that they are consistent within one sigma.
They find that the CDER technique provides improvements on the lattice with their final choices of $r_1\approx 0.9$~fm and $r_2\approx 1.3$~fm, and such improvements are insensitive to the lattice definition of operators and the HYP smearing steps within their uncertainties. 
In our work, instead of using the CDER radius cutoffs from Ref.~\cite{RBC:2014ntl}, we use 16 $L_c^4$ truncated lattices to calculate the NPR factor $Z_{O_g}^{\overline{\text{MS}}}(\mu^2,\mu^2_R)$ for each lattice spacing, which means using a 4-D cubic cutoff instead of a spherical cutoff and $L_c\approx2r_1$ and $2r_2$. 
The details of the number of measurements for each lattice spacing and $L_c$  
can be found in Table~\ref{table-Lc}.

\begin{table}[!htbp]
\centering
\begin{tabular}{|c|c|c|c|}
\hline
  ensemble & a09m310 & a12m310 & a15m310 \\
\hline
  $L_c$ & $\{8,12,16,20,24\}$ & $\{8,12,16,20\}$  & $\{8,10,12\}$ \\
\hline
  $N_\text{cfg}$ & 347  & 409  & 394 \\
\hline
  $N_\text{meas}$ & 5552  & 6544  & 6304 \\
\hline
\end{tabular}
\caption{ 
The truncation length $L_c$ in lattice units and the number of configurations $N_\text{cfg}$ and measurements $N_\text{meas}$ used for different lattice-spacing ensembles.
We used 16 sources for the truncation on each configuration;
thus, $N_\text{meas}$ is $16 \times N_\text{cfg}$.
}
\label{table-Lc}
\end{table}

The smallest cutoffs $L_c$ we use are 8 lattice units, which correspond to $0.72$, $0.96$, and $1.2$~fm for the a09m310, a12m310, and a15m310 ensembles respectively; this corresponds to $2r_1$ with similar smallest cutoff $\approx 0.8$~fm used in Ref.~\cite{Yang:2018bft}. 
Figure~\ref{fig:Z-L-dep} shows the $(Z_{O_g}^{\overline{\text{MS}}}(\mu^2=4 \text{ GeV}^2,p^2))^{-1}$ as a function of $p^2$ for different cutoffs $L_c$ for three ensembles (also the full lattices in grey points).
The error of  $Z_{O_g}^{\overline{\text{MS}}}(\mu^2=4 \text{ GeV}^2,p^2)$ becomes smaller as $L_c$ decreases, which is expected as per the $\chi$QCD results~\cite{Yang:2018bft}.
Different $L_c$ results are consistent within a one sigma error range except for the $L_c=8$ in a09m310 ensemble, likely suffering from finite-volume effects.
Our final choice of the cutoffs are $L_c=L/2\approx\{1.44, 1,44, 1.2\}$~fm for a09m310, a12m310, and a15m310 ensembles respectively where $L$ is the full lattice size.
These cutoff lengths of $L_c\in[1.2,1.44]$~fm which correspond to $r_1\approx 0.7$~fm are shown to be consistent with the full lattice NPR factors in the $\chi$QCD work~\cite{Yang:2018bft}.

\begin{figure}[htbp]
  \centering
  \includegraphics[width=0.45\textwidth]{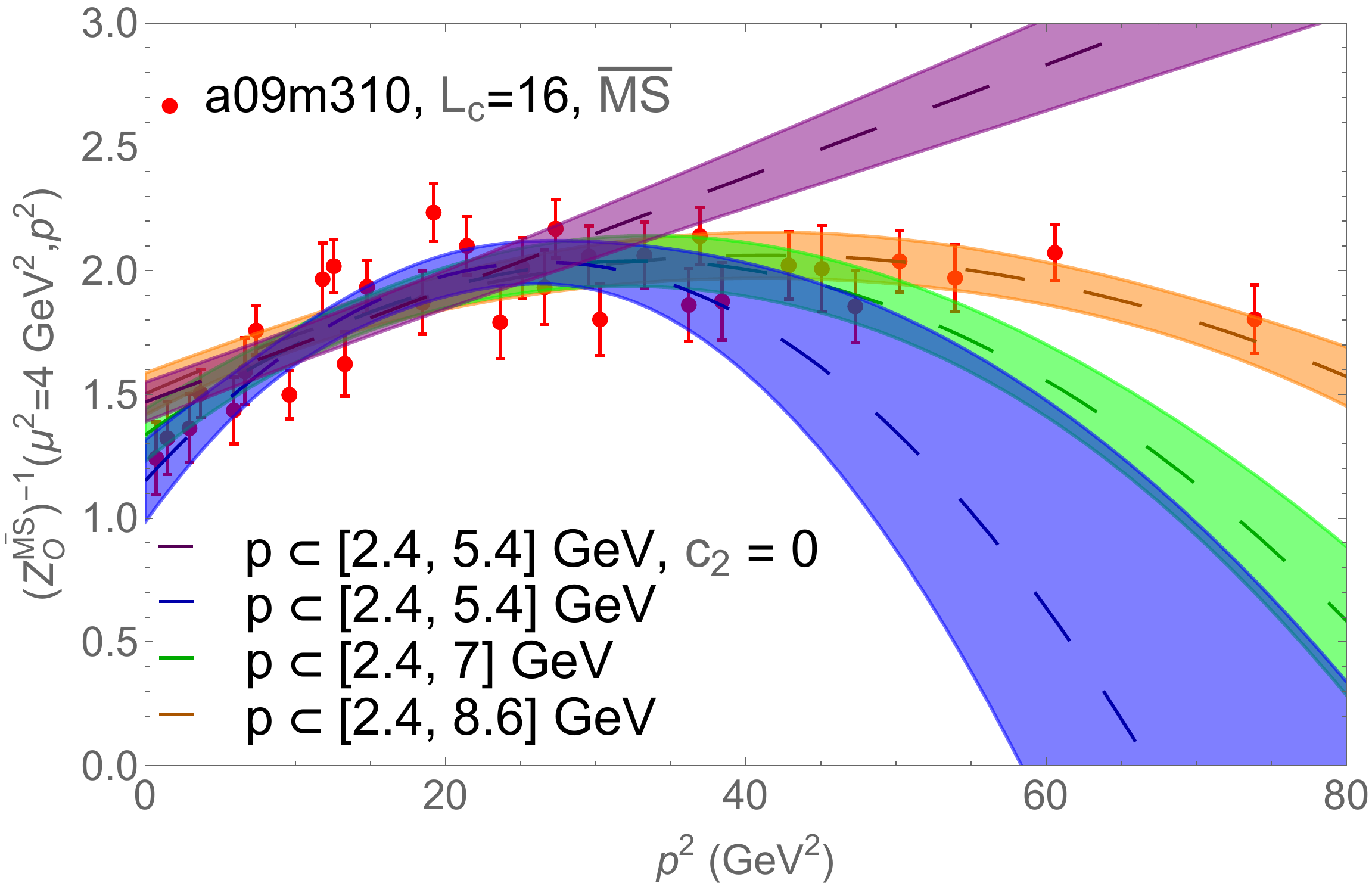}
  \includegraphics[width=0.45\textwidth]{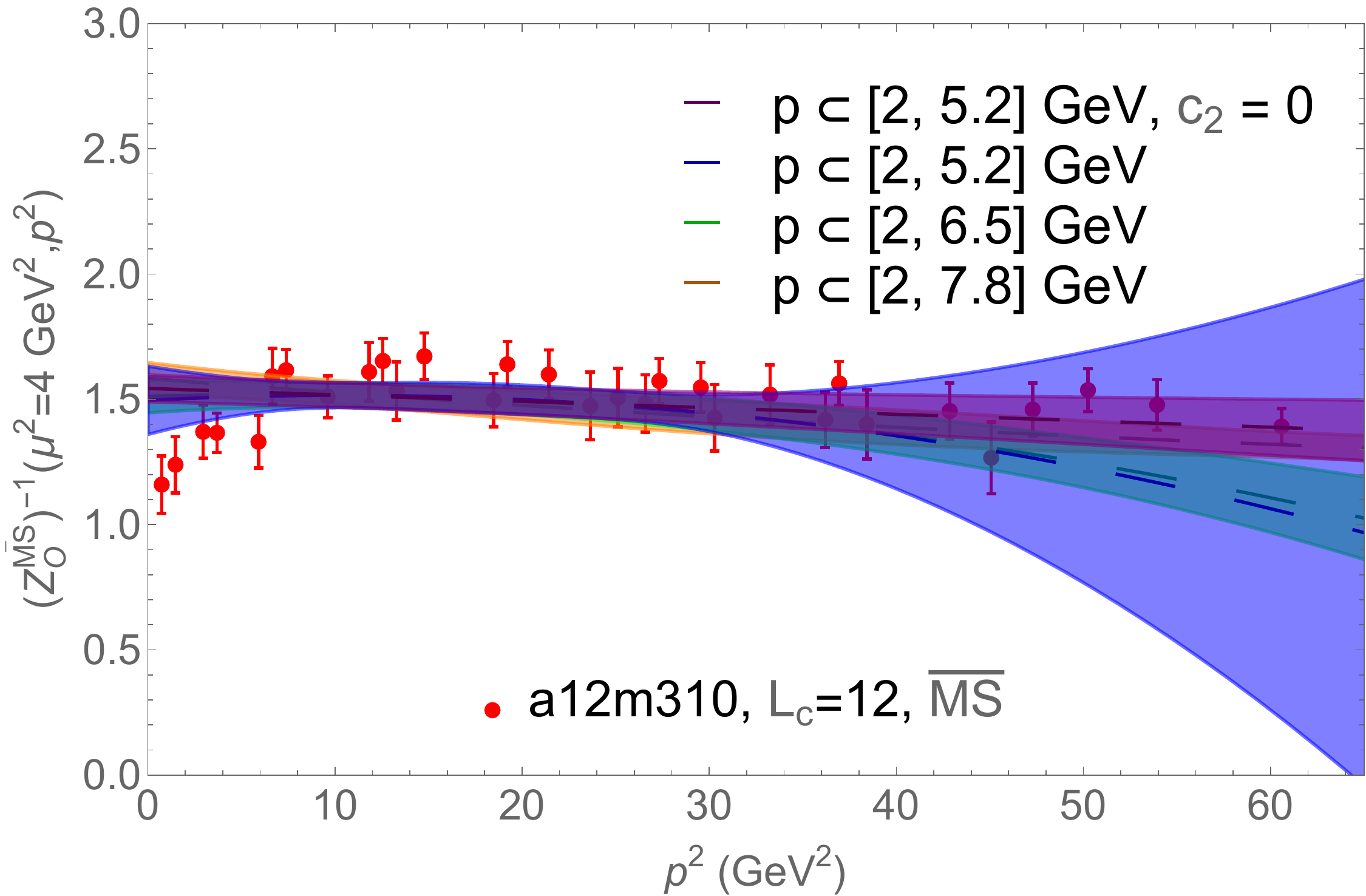}
  \includegraphics[width=0.45\textwidth]{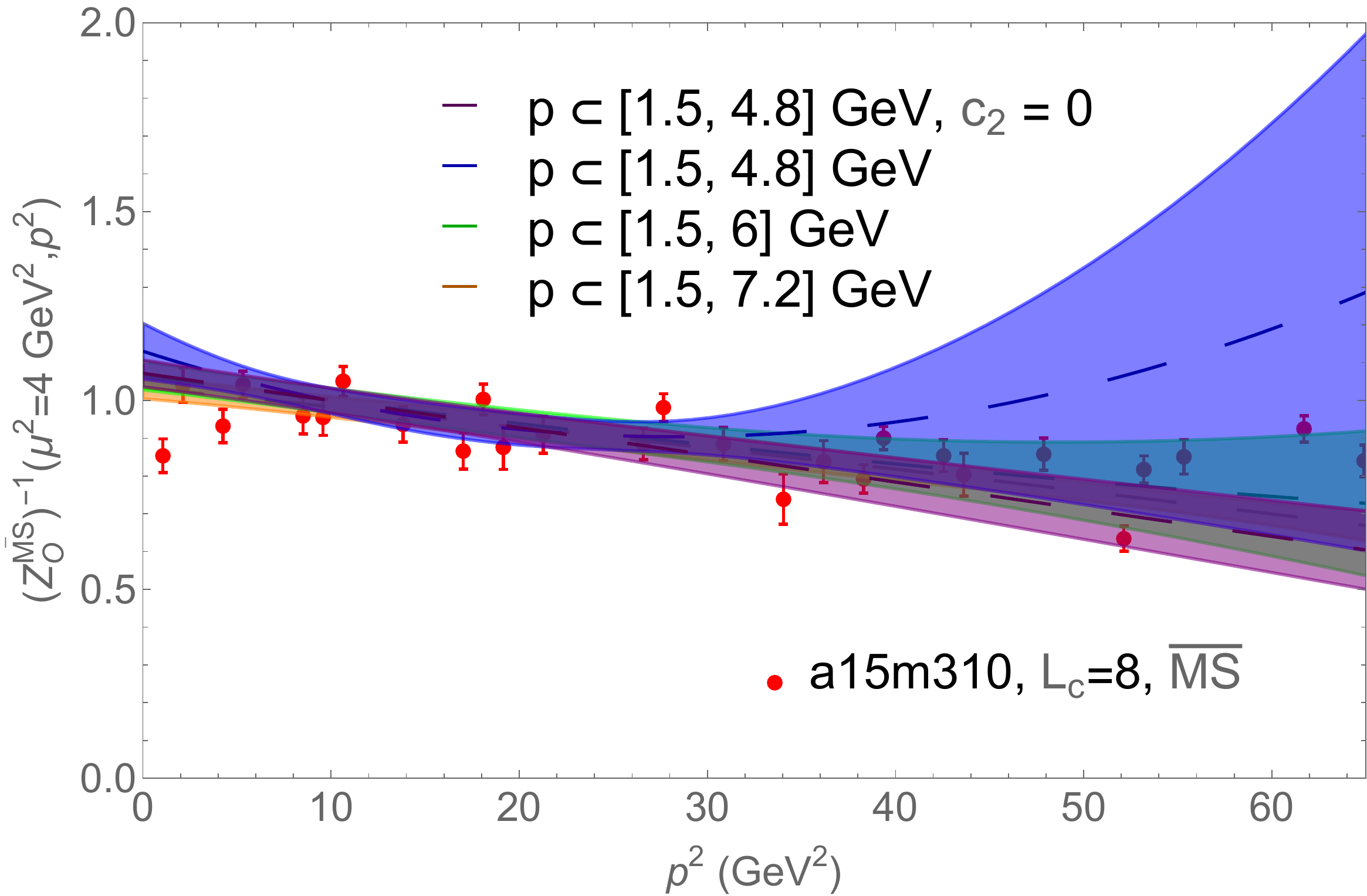}
  \caption{
  The renormalization constants $\left(Z_{O_g}^{\overline{\text{MS}}}\right)^{-1}(\mu^2=4\text{ GeV}^2,p^2)$ as a function of $p^2$ for the a09m310 $L_c=16$, a12m310 $L_c=12$, and a15m310 $L_c=8$ with various fit momentum ranges are shown from left to right respectively.
  The lower limits of the fit range of the momentum are chosen to be the same as in Ref.~\cite{Mondal:2020cmt}.
   }\label{fig:Z-p2fit}
\end{figure}

We fit the $p^2$-dependent renormalization factor 
according to the following functional form
\begin{equation}
\left(Z_{O_g}^{\overline{\text{MS}}}\right)^{-1}(\mu^2=4 \text{ GeV}^2,p^2) = \left(Z_{O_g}^{\overline{\text{MS}}}\right)^{-1} + c_1p^2 + c_2p^4,
\label{eq:Z_p2fit}
\end{equation}
where $\left(Z_{O_g}^{\overline{\text{MS}}}\right)^{-1}$ in the right hand side of the equation is the the renormalization factor at $\mu^2=4 \text{ GeV}^2$ and $p^2=0$. Figure~\ref{fig:Z-p2fit} shows examples with $L_c = L/2$ for all 3 lattice spacing ensembles and the corresponding fit bands using Eq.~\ref{eq:Z_p2fit} with ($c_2 \neq 0$) and without ($c_2=0$) the quadratic term  for large and small $p^2$ ranges used in the fit.
We only use $p$ larger than 1.5, 2.0, 2.4 GeV for $a\approx 0.15, 0.12, 0.09$~fm ensembles based on the $p_\text{min}$ used in quark momentum fractions on the same mixed-action study by PNDME~\cite{Mondal:2020cmt} (but PNDME only used constant fits to determine the renormalization constants). 
For the largest lattice spacing (a15m310 ensemble), the renormalization constants $\left(Z_{O_g}^{\overline{\text{MS}}}\right)^{-1}(\mu^2=4 \text{ GeV}^2,p^2)$ are quite linear as a function of $p^2$. Therefore, different fit bands are consistent for different fit ranges of $p^2$ with ($c_2\neq0$) and without the ($c_2=0$) the quadratic term.
The fit bands of $\left(Z_{O_g}^{\overline{\text{MS}}}\right)^{-1}(\mu^2=4 \text{ GeV}^2,p^2)$ of the a12m310 ensemble are still consistent with each other within the one-sigma error despite the large error for the smallest $p^2$ range $p\in[2, 5.2]$ GeV.
The fit bands of $\left(Z_{O_g}^{\overline{\text{MS}}}\right)^{-1}(\mu^2=4 \text{ GeV}^2,p^2)$ of the a12m310 ensemble deviate at large $p^2$ because the $\left(Z_{O_g}^{\overline{\text{MS}}}\right)^{-1}(\mu^2=4 \text{ GeV}^2,p^2)$ points increase and then decrease from small to large $p^2$, which shows that $\left(Z_{O_g}^{\overline{\text{MS}}}\right)^{-1}(\mu^2=4 \text{ GeV}^2,p^2)$ is not so linear as a function of $p^2$. However, the fit results of $\left(Z_{O_g}^{\overline{\text{MS}}}\right)^{-1}(\mu^2=4 \text{ GeV}^2,p^2)$ at $p^2=0$ start to converge at ranges with larger maximum $p^2$ chosen for the fit. Thus, we can choose $p\in[2.4, 7]$ GeV as the fit range that we use in later calculations. 
We  use the quadratic fits with [1.5, 6], [2,6.5] and [2.4, 7] GeV for each $L_c$ to extract the renormalization constants.
The renormalization constants $\left(Z_{O_g}^{\overline{\text{MS}}}\right)^{-1}$ for the three ensembles are listed in Table~\ref{table-moments}.
Using Eq.~\ref{eq:xgMSbar}, we obtain the renormalized gluon momentum fraction $\langle x \rangle_g^{\overline{\text{MS}}}$ results of four ensembles for both light and strange nucleons, listed in Table~\ref{table-moments}. 

%%%%%%%%%%%%%%%%%%%%%%%%%%%%%%%%%%%%%%%%%%%%%%%%%%%%%%%%%%%%%%%%%%%%%%%%%%%%%%%%
\section{Results and Discussion}\label{sec:results}
\begin{figure*}[htbp]
\centering
\includegraphics[width=0.46\textwidth]{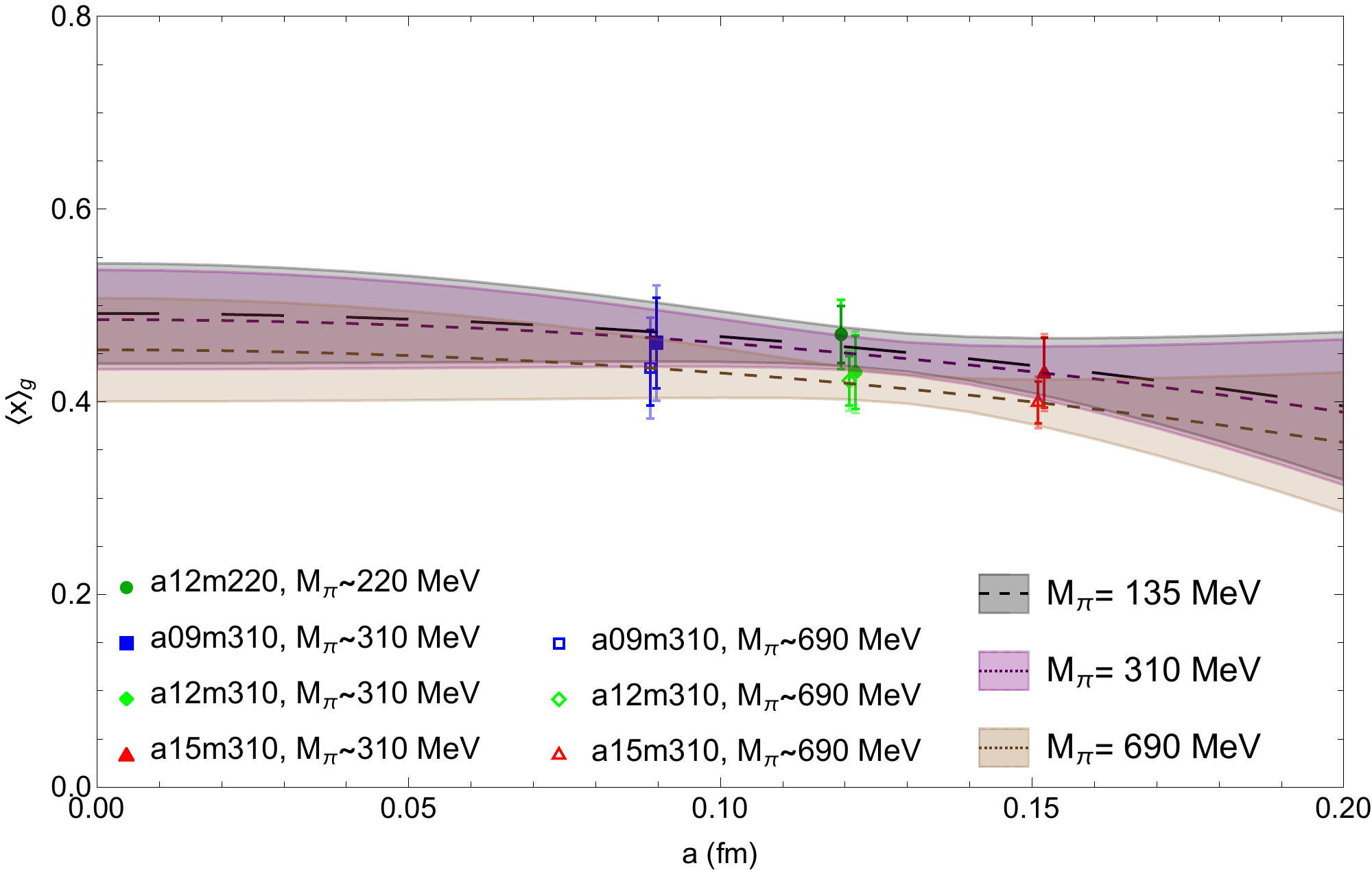}
\centering
\includegraphics[width=0.46\textwidth]{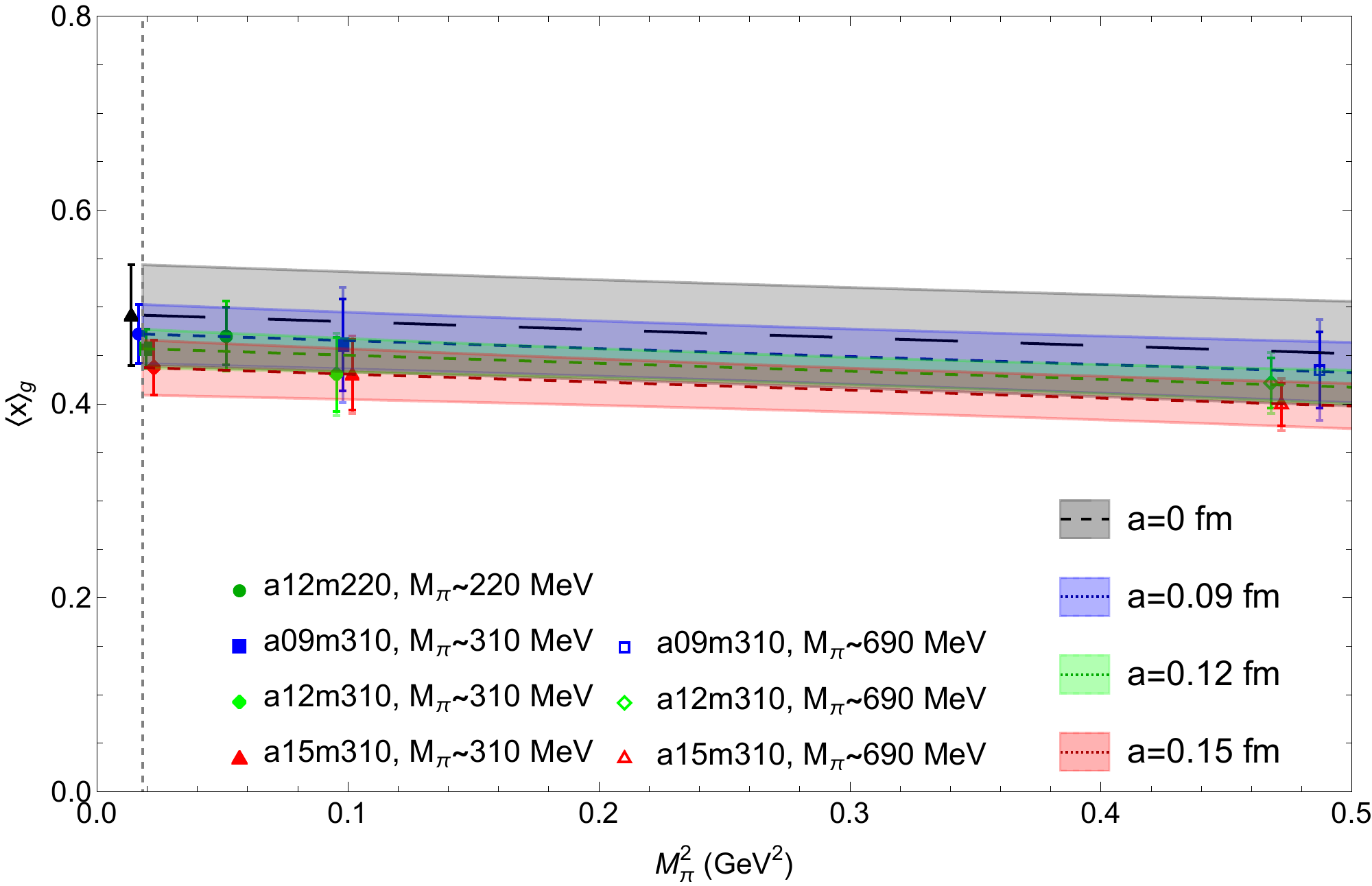}
\caption{
The renormalized gluon momentum fraction $\langle x \rangle_g^{\overline{\text{MS}}}$ obtained from each ensemble along with the physical-continuum extrapolation as functions of lattice spacing $a$ (left) and pion mass $M_\pi^2$ (right).
Each data point in the plot has two errors: the darker inner bar indicates the statistical error, while the lighter outer bar includes combined errors from both the statistical and renormalization error.
The vertical dashed line in the right plot goes through $M_\pi^2=(0.135\text{ GeV})^2$, and the different color points near this line represent the extrapolated values at different lattice spacings $a$ at physical pion mass.
To increase visibility, we plot the $M_\pi\in\{220,310\}$-MeV points shifted by $+0.001$~fm in the left plot.
The reconstructed fit bands at selected $M_\pi\in\{135, 310, 690\}$~MeV as functions of $a$ and at selected $a\in\{0, 0.09, 0.12, 0.15\}$~fm as functions of $M_\pi$ are also shown in the left- and right-side plots, respectively.}
\label{fig:xg-extra}
\end{figure*}

%%%%%%%%%%%%%%%%%%%%%%%%%%%%%%%%%%%%%%%%%%%%%%%%%%%%%%%%%%%%%%%%%%%%%%%%%%%%%%%%
Combining the results from Sections~\ref{sec:lattice-ME} and \ref{sec:NPR}, we obtain renormalized gluon momentum fractions $\langle x \rangle_g^{\overline{\text{MS}}}$ at three lattice spacings and three pion masses as shown in Fig.~\ref{fig:xg-extra}.
The points in Fig.~\ref{fig:xg-extra} have two kinds of errorbars; the darker smaller bars include only the statistical error for the gluon momentum fraction, while the lighter larger bars including both the statistical errors and the errors from the gluon NPR factor.
Our renormalized $\langle x \rangle_g^{\overline{\text{MS}}}$ shows weak pion-mass and lattice-spacing dependence.
Therefore, we use a simple quadratic ansatz for $M_\pi$ and $a$ in the physical-continuum extrapolation to the physical pion mass $M_\pi^{\text{phys}}=135$~MeV and continuum limit $a=0$:
\begin{equation}
\langle x \rangle_g^{\overline{\text{MS}}}(M_\pi,a) =
\langle x \rangle_{g}^{\overline{\text{MS}},\text{cont}}
+ k_M(M_\pi^2-(M_\pi^{\text{phys}})^2)
+ k_a a^2\ .
\end{equation}
In the fits depending on $M_\pi$ and $a$, both the statistical errors and the NPR errors are considered.
The physical-continuum limit gluon momentum fraction $\langle x \rangle_{g}^{\overline{\text{MS}},\text{cont}}$ fit result is $0.492(52)$.  
The fitted parameters $k_M=-8.2(5.0)\times 10^{-5}\text{ GeV}^{-2}$ and $k_a=-0.024(31)\text{ fm}^{-2}$ are very small, consistent with zero within two sigma.
The reconstructed fit bands at selected $M_\pi\in\{135, 310, 690\}$~MeV as functions of $a$ are shown in the left plot of Fig.~\ref{fig:xg-extra}.
There is a slight trend toward higher gluon momentum fractions as one approaches the physical pion mass.
The $M_\pi=690$~MeV band deviates from the other two bands, while the  $M_\pi=135$ and $310$~MeV bands almost coincide.
One can also see that the fit form well describes the data since these bands go through the $M_\pi=220$- and 310-MeV data points. 
On the right-hand side of Fig.~\ref{fig:xg-extra}, we show reconstructed results at $a\in\{0, 0.09, 0.12, 0.15\}$~fm as functions of $M_\pi$.
Each color band representing different lattice spacings agrees well with the same-color data points.
The central values of continuum extrapolation favor higher gluon momentum fractions but remain within one sigma of the bands from all three lattice spacings.

So far, we have been missing a systematic error associated with the mixing from the quark sector.
The bare operator in Eq.~\ref{eq:op_def} can mix with the singlet quark operators $O^\text{bare}_{q}$ and couple with the renormalized gluon operator via $O_{g} = Z_{gg}O^\text{bare}_{g} + Z_{gq}\sum_{i=u,d,s}O^\text{bare}_{q,i}$.
The mixing for quark operators is expected to be small, based on past lattice works.
The ETM Collaboration~\cite{Alexandrou:2016ekb,Alexandrou:2017oeh,Alexandrou:2020sml} used one-loop perturbative renormalization and estimated the mixing coefficients to be a fraction of their statistical errors.
The effect of the mixing of the quark operator into the gluon operator is about 2--10\%, as shown in Ref.~\cite{Alexandrou:2016ekb}.
An MIT group also ignored the quark mixing because it is assumed to be smaller than the statistical uncertainties~\cite{Shanahan:2018pib}.
We conservatively estimate a 10\% systematic error from quark mixing for this calculation;
thus, our final $\langle x \rangle_{g}^{\overline{\text{MS}},\text{cont}}$ at physical pion mass and continuum limit is  $0.492(52)_{\text{stat}+\text{NPR}}(49)_\text{mixing}$.

\begin{figure}[htbp]
\centering
\includegraphics[width=0.46\textwidth]{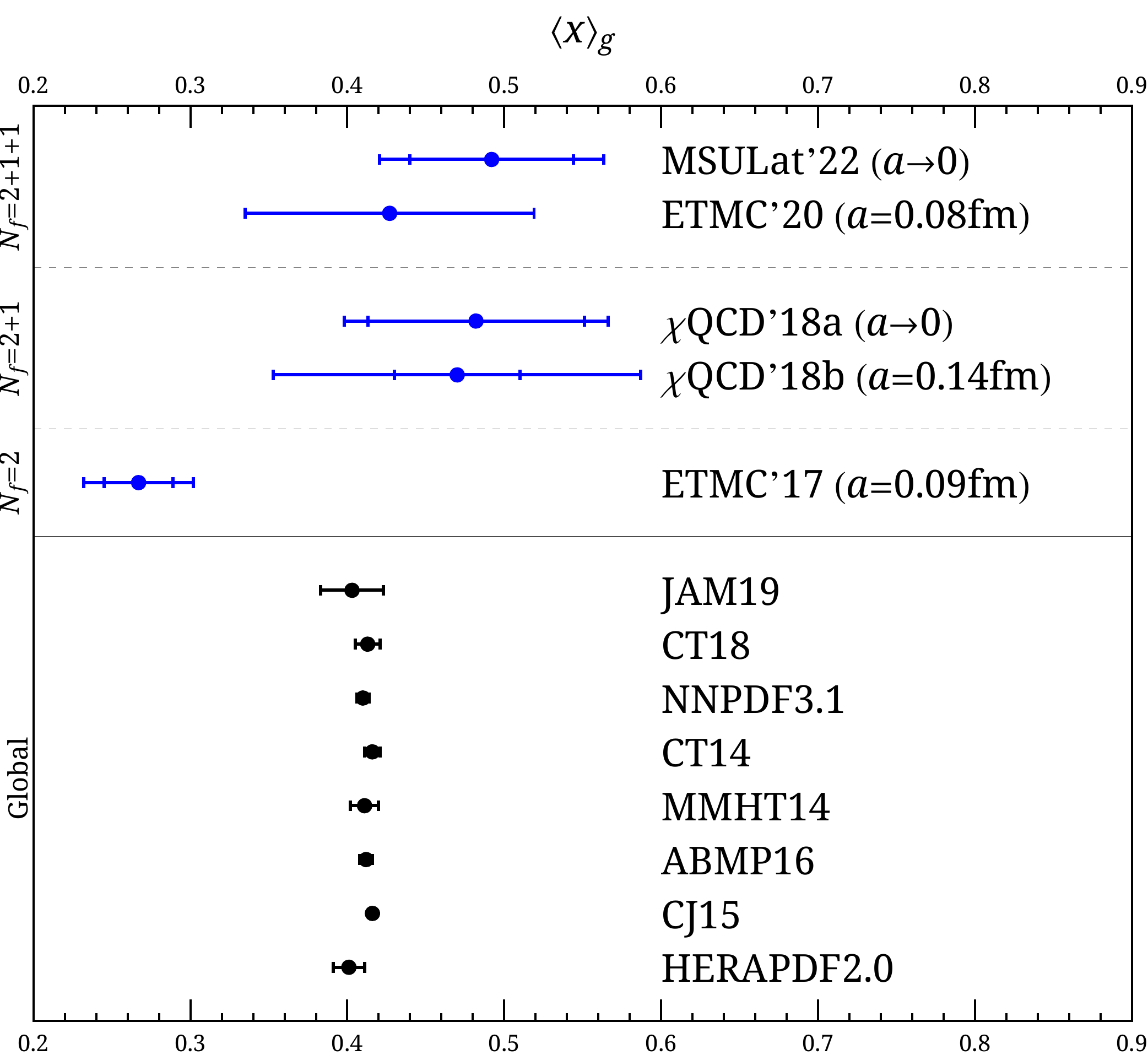}
\caption{
Comparisons of lattice-QCD and global fit determinations of the gluon moments of unpolarized PDFs at $\mu=2$~GeV.
On the lattice side, we only show those results at or extrapolated to physical pion mass by
this work (MSULat'22),
ETMC'20~\cite{Alexandrou:2020sml},
$\chi$QCD'18a\cite{Yang:2018bft},
$\chi$QCD'18b\cite{Yang:2018nqn}, and
ETMC'17~\cite{Alexandrou:2017oeh},
compared with global fit results from
JAM19~\cite{Sato:2019yez},
CT18~\cite{Hou:2019efy},
NNPDF3.1~\cite{Ball:2017nwa},
CT14~\cite{Dulat:2015mca}, MMHT14~\cite{Harland-Lang:2014zoa}, ABMP16~\cite{Alekhin:2017kpj},
CJ15~\cite{Accardi:2016qay}
and HERAPDF2.0~\cite{H1:2015ubc} analyses.
Some lattice-QCD calculations include systematic errors and some do not; we refer readers to Table~1 for more details on the difference in the errors.
Overall, the lattice calculations prefer higher central values of the gluon momentum fraction than the global fits.
}
\label{fig:xg-comp}
\end{figure}

We compare our results with prior dynamical lattice work and global fits.
As shown in Table~\ref{tab:latticemoments}, the majority of nucleon gluon momentum fractions $\langle x \rangle_g$ from lattice dynamical calculations were done using a single lattice spacing.
These results range from 0.4 to 0.55 for the most recent calculations (except the ETMC16 and ETMC17 results) and have statistical errors varying from 5--20\%.
The $\chi$QCD Collaboration studied the systematic errors from continuum extrapolation and assigned it a 10\% relative error in Ref.~\cite{Yang:2018nqn} and a 5\% relative error in their most recent paper~\cite{Wang:2021vqy}.
Overall, we find good consistency with lattice determinations from the last four years.
We summarize the dynamical lattice-QCD results extrapolated to or directly calculated at physical pion mass, along with the global-fit results since 2014, in Fig.~\ref{fig:xg-comp}.
The lattice results currently are much larger than with those from global fits, with central values closer to 0.5, rather than around 0.4, where global fits prefer.
Higher-precision lattice calculations are needed with order-of-magnitude increases in computational resources to reduce the errors to be comparable with those from global fits (using more than 60 years of experimental data).

%%%%%%%%%%%%%%%%%%%%%%%%%%%%%%%%%%%%%%%%%%%%%%%%%%%%%%%%%%%%%%%%%%%%%%%%%%%%%%%%
\section{Conclusion and Outlook}\label{sec:conclusion}
%%%%%%%%%%%%%%%%%%%%%%%%%%%%%%%%%%%%%%%%%%%%%%%%%%%%%%%%%%%%%%%%%%%%%%%%%%%%%%%%
We present the first $N_f=2+1+1$ continuum-limit lattice calculation of the gluon momentum fraction.
We use high-statistics nucleon two-point correlators ranging from 0.26--1.5~million measurements with three lattice spacings and the lightest pion mass being 220~MeV.
We apply a two-state fit to multiple source-sink separations to extract ground-state matrix elements.
We nonperturbatively calculate renormalization factors for these operators in the RI/MOM scheme, following the traditional NPR approach.
For the ensembles at pion mass 310 MeV, even though the spatial volumes are roughly the same among our three lattice spacings, the finest lattice spacing, $a\approx 0.09$~fm, yields much noisier results.
To improve this, we apply cluster-decomposition error reduction (CDER).
The renormalized gluon momentum fractions show mild lattice-spacing and pion-mass dependence (within our statistical and NPR errors);
thus, we use a simple ansatz to extrapolate to the physical-continuum limit.
Our final gluon momentum fraction is $0.492(52)_\text{stat.+NPR}(49)_\text{mixing}$, where the mixing systematic is estimated from upper bounds determined in previous lattice work.
Our lattice results are consistent with lattice work from the last four years using single lattice spacings and $N_f=2+1$ mixed action, and they are consistent with those from global fits within two sigma.
Future calculations will include ensembles at the physical pion mass and lattice calculations of the quark moments.

%%%%%%%%%%%%%%%%%%%%%%%%%%%%%%%%%%%%%%%%%%%%%%%%%%%%%%%%%%%%%%%%%%%%%%%%%%%%%%%%
\section*{Acknowledgments}

We thank MILC Collaboration for sharing the lattices used to perform this study. The LQCD calculations were performed using the Chroma software suite~\cite{Edwards:2004sx}.
%Computing resources
This research used resources of the National Energy Research Scientific Computing Center, a DOE Office of Science User Facility supported by the Office of Science of the U.S. Department of Energy under Contract No. DE-AC02-05CH11231 through ERCAP;
facilities of the USQCD Collaboration are funded by the Office of Science of the U.S. Department of Energy,
and supported in part by Michigan State University through computational resources provided by the Institute for Cyber-Enabled Research (iCER).
% Grant funding for the project
The work of  ZF and HL is partially supported by the US National Science Foundation under grant PHY 1653405 ``CAREER: Constraining Parton Distribution Functions for New-Physics Searches'' and by the  Research  Corporation  for  Science  Advancement through the Cottrell Scholar Award.

%%%%%%%%%%%%%%%%%%%%%%%%%%%%%%%%%%%%%%%%%%%%%%%%%%%%%%%%%%%%%%%%%%%%%%%%%%%%%%%%

%\bibliographystyle{unsrt}
%\bibliography{ref}

\end{document}